\documentclass[journal]{IEEEtran}

\usepackage{graphicx}
\usepackage{epstopdf}
\usepackage{standalone}
\usepackage[nolist]{acronym}
\usepackage{tcolorbox}
\usepackage{pifont}

\usepackage{pdfpages}
\usepackage{tikz}
\usetikzlibrary{positioning}
\usepackage{hyperref}
\usepackage{pdfcomment}
\usepackage{hologo}

\usepackage{xstring}

\usepackage[utf8]{inputenc}
\usepackage{marvosym}

\usepackage{algorithm}
\usepackage{algpseudocode}
\usepackage{tikz}

\usepackage{listings}
\definecolor{ListingBackground}{rgb}{0.97,0.97,0.97}
\usepackage{pgfplots}
\pgfplotsset{compat=newest}
\pgfplotsset{
	box plot/.style={
		/pgfplots/.cd,
		fill=blue!30,
		only marks,
		mark=-,
		mark size=0.2em,
		/pgfplots/error bars/.cd,
		y dir=plus,
		y explicit,
	},
	box plot box/.style={
		/pgfplots/error bars/draw error bar/.code 2 args={%
			\draw  ##1 -- ++(.2em,0pt) |- ##2 -- ++(-.2em,0pt) |- ##1 -- cycle;
		},
		/pgfplots/table/.cd,
		y index=2,
		y error expr={\thisrowno{3}-\thisrowno{2}},
		/pgfplots/box plot
	},
	box plot top whisker/.style={
		/pgfplots/error bars/draw error bar/.code 2 args={%
			\pgfkeysgetvalue{/pgfplots/error bars/error mark}%
			{\pgfplotserrorbarsmark}%
			\pgfkeysgetvalue{/pgfplots/error bars/error mark options}%
			{\pgfplotserrorbarsmarkopts}%
			\path ##1 -- ##2;
		},
		/pgfplots/table/.cd,
		y index=4,
		y error expr={\thisrowno{2}-\thisrowno{4}},
		/pgfplots/box plot
	},
	box plot bottom whisker/.style={
		/pgfplots/error bars/draw error bar/.code 2 args={%
			\pgfkeysgetvalue{/pgfplots/error bars/error mark}%
			{\pgfplotserrorbarsmark}%
			\pgfkeysgetvalue{/pgfplots/error bars/error mark options}%
			{\pgfplotserrorbarsmarkopts}%
			\path ##1 -- ##2;
		},
		/pgfplots/table/.cd,
		y index=5,
		y error expr={\thisrowno{3}-\thisrowno{5}},
		/pgfplots/box plot
	},
	box plot median/.style={
		/pgfplots/box plot
	},
	boxplot/every median/.style={
		ultra thick,dashed,cyan
	}
}

\definecolor{flexicolor}{RGB}{46,49,146}
\definecolor{amaricolor}{RGB}{237,28,36}

\usepackage{xspace}

\usepackage[binary-units=true]{siunitx}
\usepackage{psfrag}
\usepackage{graphicx}
\usepackage{tabularx,booktabs}
\usepackage{multirow, makecell}
\usepackage{xcolor,colortbl}
\usepackage{rotating}

\renewcommand{\baselinestretch}{1}

\usepackage{multirow}

\usepackage[cmex10]{amsmath}
\usepackage{amssymb}
\usepackage{wasysym}
\usepackage[caption=false,font=footnotesize]{subfig}
\usepackage{stfloats}

\begin{document}

\newcommand{\paperTitle}{The Channel as a Traffic Sensor: Vehicle Detection and Classification based on Radio Fingerprinting}

\newcommand{\paperAuthors}{Benjamin Sliwa and Christian Wietfeld}
\newcommand{\paperEmails}{$\{$Benjamin.Sliwa, Christian.Wietfeld$\}$@tu-dortmund.de}

\newcommand{\figurePadding}{0pt}
\newcommand{\figureTopPadding}{\figurePadding}
\newcommand{\figureBottomPadding}{\figurePadding}

\newcommand\attStart[1]{t_{\text{start}}(#1)}
\newcommand\attEnd[1]{t_{\text{end}}(#1)}
\newcommand\thStart{\Theta_{\text{start}}}
\newcommand\thEnd{\Theta_{\text{end}}}
\newcommand\thGuard{\Theta_{\text{guard}}}
\newcommand\durAtt[1]{\tau(#1)}

\newcommand{\msp}{\texttt{MSP430}\xspace}
\newcommand{\atmega}{\texttt{ATmega328}\xspace}
\newcommand{\esp}{\texttt{ESP32}\xspace}

\newcommand{\sfw}{0.24}

\newcommand{\dummy}[3]
{
	\begin{figure}[b!]  
		\begin{tikzpicture}
		\node[draw,minimum height=6cm,minimum width=\columnwidth]{\LARGE #1};
		\end{tikzpicture}
		\caption{#2}
		\label{#3}
	\end{figure}
}

\newcommand{\wDummy}[3]
{
	\begin{figure*}[b!]  
		\begin{tikzpicture}
		\node[draw,minimum height=6cm,minimum width=\textwidth]{\LARGE #1};
		\end{tikzpicture}
		\caption{#2}
		\label{#3}
	\end{figure*}
}

\newcommand{\basicFig}[7]
{
	\begin{figure}[#1]  	
		\vspace{#6}
		\centering		  
		\includegraphics[width=#7\columnwidth]{#2}
		\caption{#3}
		\label{#4}
		\vspace{#5}	
	\end{figure}
}
\newcommand{\fig}[4]{\basicFig{#1}{#2}{#3}{#4}{0cm}{0cm}{1}}

\newcommand{\subfig}[3]%
{%
	\subfloat[#3]%
	{%
		\includegraphics[width=#2\textwidth]{#1}%
	}%
	\hfill%
}%

\newcommand{\subfigh}[3]%
{%
	\subfloat[#3]%
	{%
		\includegraphics[height=3.65cm]{#1}%
	}%
	\hfill%
}

\newcommand\circled[1] 
{
	\tikz[baseline=(char.base)]
	{
		\node[shape=circle,draw,fill=lightgray,inner sep=1pt] (char) {\scalebox{0.66}{\textbf{#1}}};
	}\xspace
}

\newcommand\circledTab[1] 
{
	\tikz[baseline=(char.base)]
	{
		\node[shape=circle,draw,fill=lightgray,inner sep=1pt] (char) {\scalebox{0.66}{\textbf{#1}}};
	}\xspace
}

\begin{acronym}
	\acro{LIMITS}{LIghtweight Machine learning for IoT Systems}
	\acro{ITS}{Intelligent Transportation System}
	\acro{WDWS}{Wireless Detection and Warning System}
	\acro{FHWA}{Federal Highway Administration}
	\acro{NorSIKT}{Nordic System for Intelligent Classification of vehicles}
	\acro{CPS}{Cyber-physical System}
	\acro{RSSI}{Received Signal Strength Indicator}
	\acro{GPS}{Global Positioning System}
	\acro{WIM}{Weigh in Motion}
	\acro{RTI}{Radio Tomographic Imaging}
	\acro{WSN}{Wireless Sensor Network}
	\acro{CSI}{Channel State Information}
	\acro{PIR}{Passive Infrared Sensor}
	\acro{RADAR}{Radio Detection And Ranging}
	\acro{LIDAR}{Light Detection And Ranging}
	\acro{ILD}{Inductive Loop Detector}
	\acro{UAV}{Unmanned Aerial Vehicle}
	\acro{LTE}{Long Term Evolution}
	\acro{SDR}{Software Defined Radio}
	\acro{UE}{User Equipment}
	\acro{IMU}{Inertial Measuring Unit}
	\acro{IoT}{Internet of Things}
	\acro{AMR}{Anisotropic Magnetoresistive}
	\acro{MLS}{Mobile Laser Scanning}
	\acro{Euro NCAP}{European New Car Assessment Programme}
	\acro{mmWave}{millimeter Wave}
	\acro{SNR}{Signal-to-noise Ratio}
	\acro{MCU}{Microcontroller Unit}
	\acro{CPU}{Central Processing Unit}
	\acro{RAM}{Random Access Memory}
	\acro{IDE}{Integrated Development Environment}
	
	\acro{RF}{Random Forest}
	\acro{DL}{Deep Learning}
	\acro{SVM}{Support Vector Machine}
	\acro{DBT}{Deep Boltzmann Tree}
	\acro{PF}{Proximity Forest}
	\acro{RBF}{Radial Basis Function}
	\acro{ANN}{Artificial Neural Network}
	\acro{CNN}{Convolutional Neural Network}
	\acro{KNN}{k-Nearest Neighbor}
	\acro{CSR}{Classification Success Ratio}
	\acro{CART}{Classification and Regression Tree}
	\acro{WEKA}{Waikato Environment for Knowledge Analysis}
	\acro{RBF}{Radial Basis Function}
	\acro{DBM}{Deep Boltzmann Machine}
	
\end{acronym}

\acresetall
\title{\paperTitle}

\author{Benjamin Sliwa$^{1}$, Nico Piatkowski$^{2}$, and Christian Wietfeld$^{1}$%
	\thanks{$^{1}$Benjamin Sliwa and Christian Wietfeld are with Communication Networks Institute, TU Dortmund University, 44227 Dortmund, Germany
		{\tt\small $\{$Benjamin.Sliwa, Christian.Wietfeld$\}$@tu-dortmund.de}}%
	\thanks{$^{2}$Nico Piatkowski is with the Media Engineering Group, Fraunhofer Institute for Intelligent Analysis and Information Systems IAIS, 53757 Sankt Augustin, Germany
		{\tt\small $\{$Nico.Piatkowski$\}$@iais.fraunhofer.de}}
}

\maketitle
\begin{abstract}
	
%
%
Ubiquitously deployed \ac{IoT}-based automatic vehicle classification systems will catalyze data-driven traffic flow optimization in future smart cities and will transform the road infrastructure itself into a dynamically sensing \ac{CPS}.
%
%
Although a wide range of different traffic sensing systems has been proposed, the existing solutions are not yet able to simultaneously satisfy the multitude of requirements, e.g., accuracy, robustness, cost-efficiency, and privacy preservation.
%
%
In this paper, we present a novel approach, which exploits radio fingerprints -- multidimensional attenuation patterns of wireless signals -- for accurate and robust vehicle detection and classification. The proposed system can be deployed in a highly cost-efficient manner as it relies on off-the-shelf embedded devices which are installed into existing delineator posts.
%
%
In a comprehensive field evaluation campaign, the performance of the radio fingerprinting-based approach is analyzed within an experimental live deployment on a German highway, where it is able to achieve a binary classification success ratio of more than 99\% and an overall accuracy of 93.83\% for a classification task with seven different classes.

\end{abstract}


\begin{tikzpicture}[remember picture, overlay]
\node[below=5mm of current page.north, text width=20cm,font=\sffamily\footnotesize,align=center] {Published in: IEEE Internet of Things Journal\\DOI: \href{http://dx.doi.org/10.1109/JIOT.2020.2983207}{10.1109/JIOT.2020.2983207}\vspace{0.3cm}\\\pdfcomment[color=yellow,icon=Note]{
		@Article\{Sliwa2020the,\\
		Author = \{Benjamin Sliwa and Niko Piatkowski and Christian Wietfeld\},\\
		Title = \{The channel as a traffic sensor: \{V\}ehicle detection and classification based on radio fingerprinting\},\\
		Journal = \{IEEE Internet of Things Journal\},\\
		Year = \{2020\},\\
		Month = \{Mar\},\\
		Doi = \{10.1109/JIOT.2020.2983207\},\\
		\}
}};
\node[above=5mm of current page.south, text width=15cm,font=\sffamily\footnotesize] {2020~IEEE. Personal use of this material is permitted. Permission from IEEE must be obtained for all other uses, including reprinting/republishing this material for advertising or promotional purposes, collecting new collected works for resale or redistribution to servers or lists, or reuse of any copyrighted component of this work in other works.};
\end{tikzpicture}

\IEEEpeerreviewmaketitle

\section{Introduction} \label{sec:introduction}		

%
%
Reliable and efficient transportation systems are one of the key foundations for emerging \ac{IoT}-enabled smart cities \cite{Zanella/etal/2014a}. However, with the expected massive increases in vehicular traffic, e.g., through introduction of novel modes in personal transportation as well as in the logistics sector, many existing traffic systems are expected to reach their capacity limits. 
%
%
Increasing the road capacity in straightforward ways -- e.g., through construction of new lanes and roads -- is often not possible due to the involved costs and spatial limitations (especially in inner city environments). Therefore, traffic flow optimization aiming to utilize the \emph{existing} infrastructure in a more efficient way is one of the catalysts for next-generation \emph{data-driven} \acp{ITS} \cite{Zhang/etal/2011a}. In addition to abstract traffic indicators such as traffic flow and traffic density, those systems exploit \emph{vehicle type information} in order to enable novel optimization methods such as type-specific lane clearance, smart parking and type-specific routing.
%
%
In order to enable these methods, up-to-date traffic indicators need to be acquired continuously and accurately by ubiquitously deployed traffic sensors. Although the cars themselves can be exploited as moving sensors \cite{Sliwa/etal/2018e, Sliwa/etal/2019a}, it is expected that the majority of the measurements in the near future will be performed by static deployments of \ac{IoT}-enabled sensor installations \cite{Sliwa/etal/2019b}.
%
%
Those systems need to fulfill different goals in parallel: In addition to providing high detection and classification accuracies in real-time, they should work reliably even in challenging traffic and weather conditions. Furthermore, they should be privacy-preserving, energy- and cost-efficient. The latter is of severe importance in order to enable large-scale deployments for smart city applications.

%
%
\fig{b}{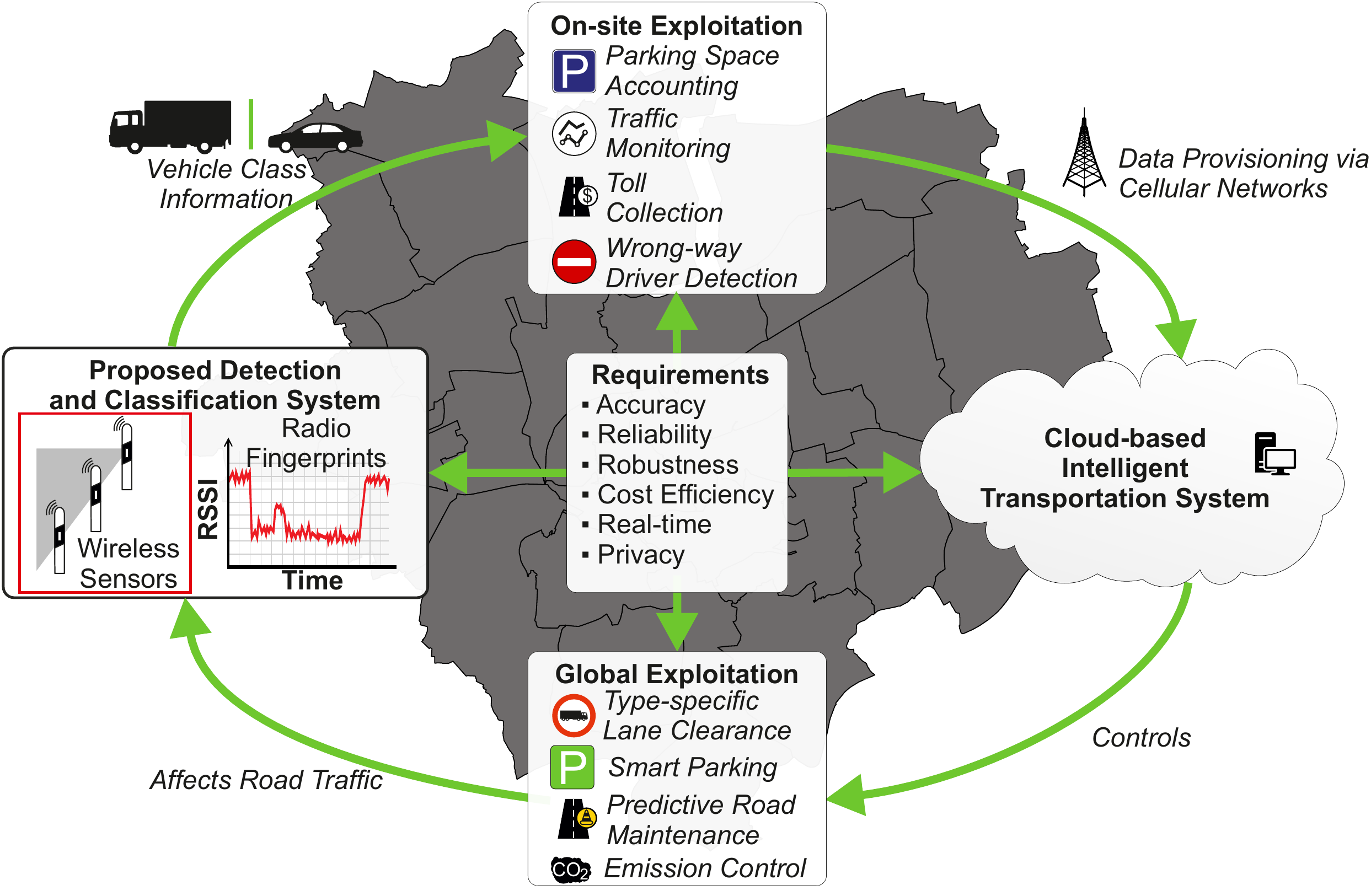}{Overall system vision: Embedding of the proposed \ac{IoT}-based sensor system in a smart city environment. All sensor deployments are locally exploited for \emph{on-site} applications and contribute their data to the \emph{global} data-driven \ac{ITS} applications.}{fig:scenario}
%
%
Although a wide range of different sensor systems has been proposed, existing approaches often have characteristic shortcomings, which limit their suitability for ubiquitous deployments.
%
%
In this paper, we present a novel vehicle detection and classification system, which is based on the \emph{radio fingerprints} of vehicles passing a \ac{WSN} installation. Although the term \emph{fingerprint} corresponds to environment-dependent characteristics of \emph{individual} vehicles, the proposed approach relies on the assumption that vehicles of the same class will have similar shapes and similar corresponding radio fingerprints. Machine learning is applied to detect those similarities between the different measurements.
Our method is inherently privacy-preserving, real-time capable and robust against challenging weather conditions. Moreover, in the considered evaluation scenario represented by a single lane deployment at a highway ramp, it is able to classify different vehicle types with a high accuracy. Moreover, as it relies on off-the-shelf hardware and is installed into the existing traffic infrastructure -- inside the housings of delineator posts -- it can be deployed highly cost-efficiently in an ad-hoc manner.
%
%
Fig.~\ref{fig:scenario} shows the intended flow of information of the proposed \ac{IoT}-based sensor system and its application in a smart city environment. The sensor deployments gather radio fingerprints of the passing vehicles and perform the classification tasks. The obtained information is used for \emph{on-site} purposes, e.g., parking space accounting, detection of wrong-way drivers, traffic monitoring and toll collection. Furthermore, the data is forwarded through a wide area communication network to a cloud-based \ac{ITS} in order to be exploited \emph{globally} for traffic flow optimization. The impact of the optimization methods affects the traffic situation, which is measured by the sensor system.
%
%
The proposed system has evolved from the \ac{WDWS} for radio-based detection of wrong-way drivers, which has been initially presented in \cite{Haendeler/etal/2014a} and patented in \cite{Wietfeld/etal/2012a}. First approaches for integrating vehicle classification capabilities using conventional data analysis methods have been presented in \cite{Haferkamp/etal/2017b, Sliwa/etal/2018e}.
The contributions provided by this paper are as follows:
\begin{itemize}
	\item Presentation of a novel vehicle detection and classification system based on \textbf{radio fingerprints}.
	\item Performance comparison of established and state-of-the-art \textbf{machine learning} methods: Achievable classification accuracy and implied resource efficiency of \acp{RF}, \acp{SVM}, \acp{DBM}, and \acp{PF}.
	\item Scalability analysis based on sub-set evaluations of the involved radio links and memory optimization of the machine learning models for different real world \acp{MCU}.
\end{itemize}
%
%
The remainder of the paper is structured as follows. After discussing related classification approaches in Sec.~\ref{sec:related_work}, we present the machine learning-based solution approach and the architecture model of the proposed system in Sec.~\ref{sec:approach}. Afterwards, the real world setup for the data acquisition is introduced in Sec.~\ref{sec:methods} and the achieved classification results are discussed in Sec.~\ref{sec:results}.

\section{Related Work} \label{sec:related_work}

In this section, an overview about existing systems and technologies for vehicle classification is provided.
%
%
\fig{b}{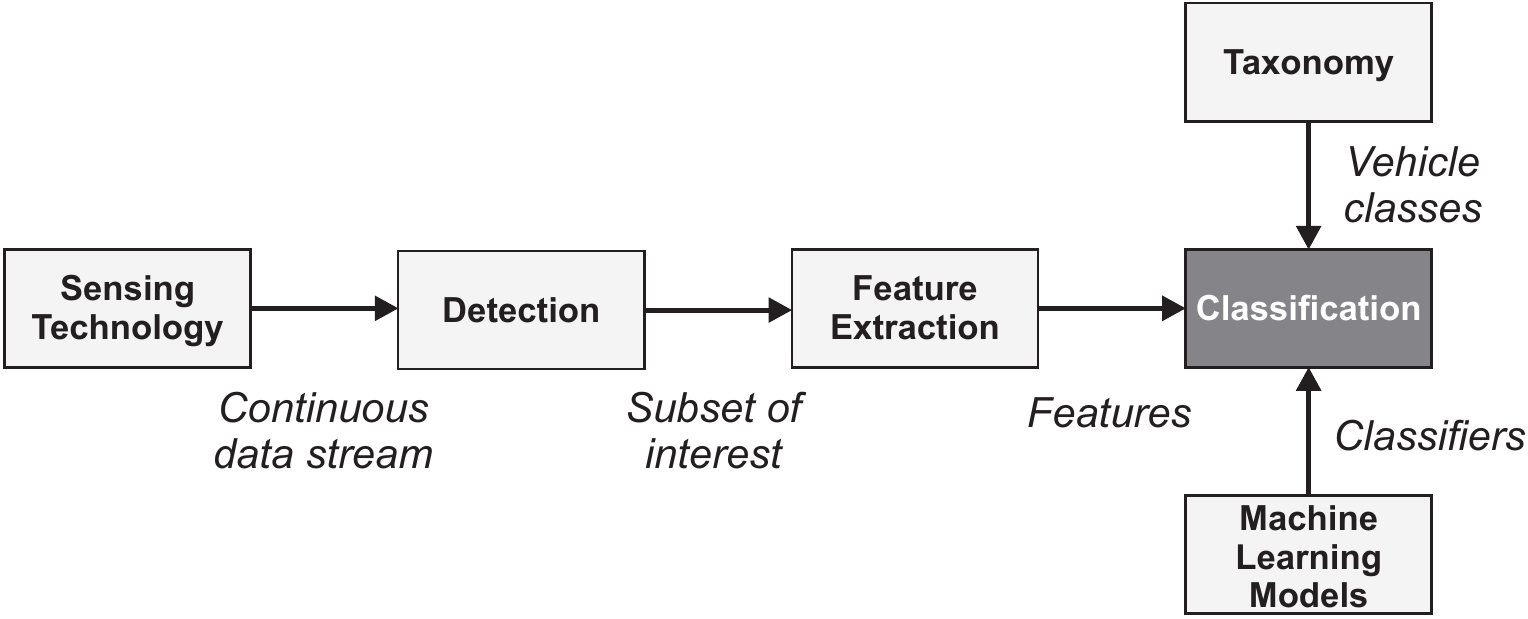}{Abstract system model for the logical information flow within vehicle classification systems.}{fig:abstract_model}
Fig.~\ref{fig:abstract_model} provides an abstract overview about the information flow and the involved logical components which ultimately lead to a vehicle classification result. A sensing technology is applied for acquiring a \emph{continuous data stream}, for which a detector extracts the \emph{subset of interest} which contains the actual measurements of the vehicle to be classified. Based on this data set, relevant \emph{features} are extracted and utilized by different \emph{classifiers} to perform the machine learning-based data analysis with respect to a well-defined \emph{classification taxonomy}.

\subsection{Taxonomies for Classification of Road Vehicles} \label{sec:taxonomies}

%
%
The \ac{FHWA} has proposed a 13-category classification scheme  \cite{Klein/etal/2006a, Transportation/2016a}, which is mostly based on the number of axles.
However, the class information itself does not allow to derive length information of individual vehicles (e.g., for parking space accounting), as multiple classes (e.g., \ac{FHWA} classes 2 and 3) optionally contain trailers.

%
%
The \ac{NorSIKT} \cite{Vaa/etal/2012a} is a hierarchical scheme with four levels of different granularity (up to 14 classes at level 4), which is applied in Sweden, Denmark, Island, Finland and Norway. 

Other popular taxonomies are provided by the \emph{{ISO 3833-1977}} scheme with 7 categories, the directive \emph{{2007/46/EG}} of the European parliament and the \ac{Euro NCAP} scheme.

However, although these standardized systems exist, most academic approaches apply their own \emph{application-specific} schemes. As an example Gebru et al. \cite{Gebru/etal/2017a} propose a hierarchical taxonomy which consists of make, model, body type and year/trim of a vehicle for visual census estimation. For the following performance evaluation of the proposed system, we apply multiple taxonomies with different grades of granularity which are further described in Sec.~\ref{sec:taxonomies}.

\subsection{Vehicle Detection and Classification Systems}

%
%
\newcommand\full{$\CIRCLE$}
\newcommand\partly{$\LEFTcircle$}
\newcommand\missing{$\Circle$}
\newcommand\ruleT{\cmidrule(ll){4-6} \cmidrule(ll){7-10} \cmidrule(ll){11-17}}

\newcommand\rot[1]{\rotatebox{90}{#1}}
\newcommand\cW{0.01cm}

\newcolumntype{A}{>{\centering\arraybackslash}m{1.8in}}
\newcolumntype{S}{>{\centering\arraybackslash}m{2in}}
\newcolumntype{N}{>{\centering\arraybackslash}m{1.0in}}
\newcolumntype{P}{>{\centering\arraybackslash}m{0.6in}}
\newcolumntype{D}{>{\centering\arraybackslash}m{0.5in}}
\newcolumntype{C}{>{\centering\arraybackslash}m{1.5in}}
\newcolumntype{G}{p{0.05 cm}}
\newcommand*\graycell{} 

\newcommand{\detection}{\circledTab{1}}
\newcommand{\otherCam}{\circledTab{2}}
\newcommand{\otherRadio}{\circledTab{3}}


\begin{table*}[ht]
	\centering
	\caption{Comparison of Existing Systems and Technologies for Vehicle Detection and Classification}
	\begin{tabular}{lp{5cm}p{0.3cm} DPD GGGG GGGGGGG}
		
		\toprule
		%
		%
		\textbf{Category} & \multicolumn{2}{S}{\textbf{System}}  & \multicolumn{3}{A}{\textbf{Accuracy}} & \multicolumn{4}{P}{\textbf{Properties}} & \multicolumn{7}{N}{\textbf{Vulnerabilities}} \\
		\ruleT
		& & & {Detection [\%]} & {Classification [\%]} & Number of Classes & \rot{Cost-efficient} & \rot{Non-intrusive} & \rot{Privacy-preserving} & \rot{Online} & \rot{\emph{Lighting}} & \rot{\emph{Temperature}} & \rot{\emph{Weather}} & \rot{\emph{Noise}} & \rot{\emph{High Speeds}} & \rot{\emph{Congestion}} & \rot{\emph{Other}} \\
		\midrule
		
		%
		%
		\multirow{1}{1cm}{\textbf{Trajectory}}
		& Acceleration analysis of \acs{GPS} traces & \cite{Simoncini/etal/2016a} & - & 95.72 & 2 & \full & \full & \full & \missing & \missing & \missing & \missing & \missing & \missing & \missing & \missing \\
		\midrule

		%
		%
		\multirow{7}{1cm}{\textbf{Intrusive}}
		& \acs{WIM} & \cite{Klein/etal/2006a} & ? & 90 & 13 & \missing & \missing & \full & \full & \missing & \missing & \missing & \missing & \partly & \missing & \missing \\
		& Dual induction loop & \cite{Wu/Coifman/2014a} & ? & 99 & 3 & \missing & \missing & \full & \full & \missing & \missing & \missing & \missing & \missing & \partly & \missing \\
		& Single induction loop & \cite{Coifman/Kim/2009a} & ? & 76.4-99 & 3 & \missing & \missing & \full & \full & \missing & \missing & \missing & \missing & \missing & \partly & \missing \\
		& Single induction loop with spectral features & \cite{Lamas-Seco/etal/2015a} & ? & 94.72-96.44 & 3 & \missing & \missing & \full & \full & \missing & \missing & \missing & \missing & \missing & \partly & \missing \\
		& Fiber Bragg grating sensors & \cite{Al-Tarawneh/etal/2018a} & ? & 98.5 & 3 & \missing & \missing & \full & \full & \missing & \missing & \missing & \missing & \missing & \partly & \missing \\
		& In-pavement vibration sensors & \cite{Ye/etal/2019a} & ? & 91.99 & 10 & \missing & \missing & \full & \full & \missing & \missing & \missing & \missing & \missing & \partly & \missing \\
		& Single piezoelectric sensor & \cite{Rajab/etal/2016a} & ? & 88.33-97.35 & 10 & \missing & \missing & \full & \full & \missing & \partly & \missing & \missing & \partly & \missing & \missing \\
		\midrule
		
		%
		%
		\multirow{4}{1cm}{\textbf{Acoustics}}
		& Pure acoustic signal analysis & \cite{George/etal/2013a} & ? &  73.42 & 3 & \full & \full & \full & \full & \missing & \missing & \partly & \full & \partly & \partly & \missing \\
		& Audio visual cues & \cite{Daniel/Mary/2016a} & \detection & 92.67 & 7 & \full & \full & \full & \full & \missing & \missing & \partly & \full & \partly & \partly & \missing \\
		& Audio visual cues & \cite{Piyush/etal/2016a} & \detection & 82 & 7 & \full & \full & \full & \full & \missing & \missing & \partly & \full & \partly & \partly & \missing \\
		& Combination with inertial sensors & \cite{Kerekes/etal/2017a} & \detection & 94 & 7 & \full & \full & \full & \full & \missing & \missing & \partly & \full & \missing & \missing & \missing \\
		\midrule

		%
		%
		\multirow{5}{1cm}{\textbf{Inertial}}
		& Accelerometers and magnetometers  & \cite{Ma/etal/2014a} & 99.19 & 98.67 & 3 & \missing & \missing & \full & \full & \missing & \partly & \missing & \missing & \missing & \missing & \missing \\
		& Accelerometers and magnetometers & \cite{Kleyko/etal/2015a} & ? & 93.4 & 3 & ? & ? & \full & \full & \missing & \partly & \missing & \missing & \missing & \missing & \missing \\
		& Wireless magnetometer & \cite{Balid/etal/2018a} & 99.98 & 97 & 4 & \full & \full & \full & \full & \missing & \partly & \missing & \missing & \missing & \missing & \missing \\
		& Magnetometer at road side & \cite{Xu/etal/2018a} & ? & 95.46 & 4 & \full & \full & \full & \full & \missing & \partly & \missing & \missing & \missing & \missing & \missing \\
		& Magnetoresistive sensor & \cite{Yang/Lei/2015a} & 99.05 & 93.66 & 5 & \full & \full & \full & \full & \missing & \partly & \missing & \missing & \missing & \missing & \missing \\
		\midrule
		
		%
		%
		\multirow{8}{1cm}{\textbf{Vision}}
		& Camera (Vehicle make and model) & \cite{Siddiqui/etal/2015a} & ? & 95.77 & 29 & \partly & \full & \missing & \full & \full & \missing & \full & \missing & \missing & \partly & \otherCam \\
		& Camera-based regression analysis & \cite{Liang/etal/2015a} & ? & 63.4-92.7 & 3 & \partly & \full & \missing & \full & \full & \missing & \full & \missing & \missing & \partly & \otherCam \\
		& Camera & \cite{Dong/etal/2015a} & ? & 88.11-95.7 & 6 & \partly & \full & \missing & \full & \full & \missing & \full & \missing & \missing & \partly & \otherCam \\
		& EasiSee: Camera + Magnetometer & \cite{Wang/etal/2014a} & 95.31 & 93 & 3 & \partly & \full & \missing & \full & \full & \missing & \full & \missing & \missing & \partly & \missing \\
		& \acs{UAV}-based aerial imaging & \cite{Liu/Mattyus/2015a} & 78.99 & 98.2 & 2 & \partly & \full & \missing & \full & \full & \missing & \full & \missing & \missing & \missing & \missing \\
		& \acs{PIR} & \cite{Odat/etal/2018a} & 99.99 & 99 & 5 & \partly & \full & \full & \full & \missing & \missing & \full & \missing & \partly & \missing & \missing \\
		& \acs{LIDAR} & \cite{Lee/Coifman/2015a} & ? & 99.5 & 3 & \partly & \full & \full & \full & \missing & \missing & \full & \missing & \missing & \missing & \missing \\
		& Laser scanning & \cite{Xiao/etal/2016a} & 95.9 & 86 & 4 & \partly & \full & \full & \full & \missing & \missing & \full & \missing & \partly & \missing & \missing \\
		\midrule
		
		%
		%
		\multirow{7}{1cm}{\textbf{Radio}}
		& Bluetooth low energy beaconing & \cite{Bernas/etal/2018a} & 98 & 97 & 3 & \full & \full & \full & \full & \missing & \missing & \missing & \missing & \missing & \missing & \otherRadio \\
		& WiFi \ac{CSI} & \cite{Won/etal/2018a} & 99.4 & 91.1 & 5 & \full & \full & \full & \full & \missing & \missing & \missing & \missing & \missing & \missing & \otherRadio \\
		& LTE-CommSense (Vehicle identification) & \cite{Sardar/etal/2019a} & ? & 92.6 & 3 & \full & \full & \full & \full & \missing & \missing & \missing & \missing & \missing & \missing & \otherRadio \\
		& \acs{RADAR}-based height profile analysis & \cite{Urazghildiiev/etal/2007a} & ? & 85 & 5 & \partly & \full & \full & \full & \missing & \missing & \missing & \missing & \missing & \partly & \otherRadio \\
		& Radio fingerprinting & \cite{Sliwa/etal/2018e} & ? & \{99, 89.15\} & \{2, 9\}  & \full & \full & \full & \full & \missing & \missing & \missing & \missing & \missing & \missing & \otherRadio \\
		\cline{2-17} \vspace{-0.2cm} \\ 
		& \graycell \textbf{This paper} & \graycell & \graycell \textbf{100} & \graycell \textbf{ \{99.08, 95.9, 93.8\}} & \graycell \textbf{\{2, 3, 7\}} & \graycell \full & \full & \graycell \full & \graycell \full & \graycell \missing & \graycell \missing & \graycell \missing & \graycell \missing & \graycell \missing & \graycell \missing & \otherRadio \\
		\bottomrule
		
	\end{tabular}
	\label{tab:related_work}
	
	\vspace{0.1cm}
	-: Missing, ?: Unknown, \missing: Low / Does not apply, \partly: Middle / Partial, \full: High, Full, \detection: Absence of vehicles is treated a specific class \\
	\otherCam: Camera-based systems might be subject to fraud attempts such as adversarial attacks, \otherRadio: Radio-based systems in the unlicensed frequency spectrum might be affected by interference and could be attacked by radio jammers\\
\end{table*}

%
%
A summary about well established sensor systems (e.g., \acp{ILD} and \ac{WIM}) used by the \ac{FHWA} is provided by \cite{Klein/etal/2006a}. More research-oriented summaries of existing systems and open research directions are given in \cite{Won/2019a} and \cite{Guerrero-Ibanez/etal/2018a}.
%
%
Although the following literature analysis focuses on comparing the resulting classification accuracy of different sensor systems, an important prerequirement is the capability for detecting vehicles entering the area covered by the systems.
Apart from so far unusual approaches, such as the postprocessing analysis of \ac{GPS} traces \cite{Simoncini/etal/2016a}, different categories of detection and classification systems can be identified, which are analyzed in the following.
%
%
However, it needs to be noted that comparing the resulting accuracies of different systems is a non-trivial task since the number of classes and the class definitions differ among the considered research works.

%
%
The majority of the existing approaches rely on well-known machine learning methods for performing the actual classification tasks with \acp{SVM} \cite{Cortes/Vapnik/1995a}, \acp{RF} \cite{Breiman/2001a}, \ac{KNN} and different variants of (deep) \acp{ANN} \cite{Goodfellow/etal/2016a} being the most commonly used models.
%
%
Although the classification can be performed on the raw \emph{time series} data, most systems apply \emph{featured-based} mechanisms, e.g., based on estimations of the vehicle length.
Only a few authors mention the hardware platform used to execute the actual machine learning based-classification (e.g., \cite{Xu/etal/2018a} use an off-the-shelf notebook for this task). In the vast majority of the analyzed research works, the authors acquire sensor data in the real world but perform the data analysis offline (e.g., as explicitly discussed by \cite{Won/etal/2018a}). However, deployment machine learning models to resource-constrained \ac{IoT} platforms such as ultra low power microcontrollers has become an emerging research field \cite{Sliwa/etal/2020c}. Therefore, it can be expected that platform-related dimensioning of machine learning models will become a more discussed topic in the vehicle classification community in the near future. 
Although most of our performance evaluations in Sec.~\ref{sec:results} are also based on offline data analysis, we discuss and evaluate platform- and model-specific deployment aspects in Sec.~\ref{sec:platforms}.

Tab.~\ref{tab:related_work} provides a comparison table about existing solution approaches and their characteristic properties and classification accuracies. If not explicitly stated within the considered studies themselves, the system properties and vulnerability evaluations are based on the considered summary papers. As a reading example, the \ac{WIM} system of \cite{Klein/etal/2006a} does not mention detection accuracies but achieves 90\% classification accuracy on a two-class problem. \ac{WIM} systems have a low cost efficiency, as their intrusive installation requires heavy roadwork. However, the are highly privacy-preserving and can be utilized for online classification. They are robust against most considered vulnerabilities but are affected by high velocities.

%
%
\textbf{Intrusive systems} form the \emph{classic} approach for measuring traffic indicators at strategic locations on the pavement surfaces \cite{Guerrero-Ibanez/etal/2018a}. As the installation and maintenance requires heavy roadwork, the involved costs are very high, which makes those systems not suitable for large-scale deployments.
\ac{WIM} systems usually combine multiple \acp{ILD} and piezoelectric pressure sensors. Due to the multitude of involved subsystems and the required roadwork, those systems are highly expensive and are therefore only deployed at chosen locations. There is practically no active research on \ac{WIM} systems themselves, yet some research works exploit a nearby \ac{WIM} as ground truth. 
%
%
\ac{ILD}-based approaches usually rely on a dual-loop setup \cite{Wu/Coifman/2014a, Bitar/Refai/2017a} for estimating the vehicle length, which is then exploited for the vehicle classification. However, it has been shown that single-loop detectors can approach a similar level of accuracy \cite{Coifman/Kim/2009a}. However, as discussed in \cite{Ma/etal/2014a}, loop-based methods do not perform well under congested traffic.
%
%
Lamas-Seco et al. \cite{Lamas-Seco/etal/2015a} propose to exploit spectral features, which do not depend on the vehicle speed and allow to perform vehicle classification with a single-loop approach.
%
%
In \cite{Al-Tarawneh/etal/2018a}, the authors use fiber Bragg grating sensors to achieve 98.5\% accuracy for three different classes based on \ac{SVM}.
%
%
Ye et al. \cite{Ye/etal/2019a} use in-pavement vibration sensors to calculate speed and axle spacing, which is then utilized for vehicle classification based on \ac{ANN} and \ac{KNN}.
%
%
The authors of \cite{Rajab/etal/2016a} aim to reduce the deployment costs of \ac{WIM} systems by performing vehicle classification based on a single piezoelectric sensor.

%
%
\textbf{Acoustic sensor systems} rely on audio signals for vehicle detection and classification.
A key challenge is the removal of undesired noise, which is omnipresent in the considered traffic scenario and requires multiple preprocessing and filtering steps. 
%
%
However, the achievable accuracy of solely acoustics-based systems is relatively low (e.g., 73.42\% in \cite{George/etal/2013a}).
%
%
Audiovisual cues aim to compensate the shortcomings of the individual systems by combining the low computation complexity of audio systems for vehicle detection with the high classification accuracy of camera systems. Using a deep \ac{ANN}, Daniel et al. \cite{Daniel/Mary/2016a} achieve 92.67\% accuracy on seven vehicles classes, although the individual sensors only achieved 82\% (camera) and 66\% (audio).
%
%
A similar study is carried out in \cite{Piyush/etal/2016a}.
%
%
Kerekes et al. \cite{Kerekes/etal/2017a} furthermore combine acoustics, magnetometers and electromagnetic radio field sensing for identifying different vehicles.

%
%
\textbf{Inertial sensors} are small-scale and low-cost sensors, which usually measure a single indicator (e.g., the magnetic field) in multiple dimensions. In many cases, different sensors (e.g., accelerometers, barometers and magnetometers) are combined within an \ac{IMU} which is then installed either on or within the pavement or at the side of the road. The resulting deployment costs are mainly related to the installation method and not the sensors themselves. 
%
%
Many research works apply \acp{AMR} to detect the number of axles of the passing vehicle.
%
%
Ma. et al \cite{Ma/etal/2014a} use a combination of magnetometers and accelerometers, which is compared against video images and a nearby \ac{WIM} system. Although the resulting accuracy is 99\% for distinguishing between 2-axle cars, 3-axle cars and 5-axle heavy trucks, the size of the data set is relatively low.
%
%
Kleyko et al. \cite{Kleyko/etal/2015a} analyse the combined usage of magnetometers and accelerometers based on a large data set for vehicle classification according to the \ac{NorSIKT} taxonomy. They apply feature-free \emph{data smashing} for determining if two data streams were produced by the same source.
%
%
Balid et al. \cite{Balid/etal/2018a} achieves 99.98\% detection accuracy and 97\% length-based vehicle classification accuracy on a four-group scheme. Similar experiments are performed by \cite{Xu/etal/2018a} and \cite{Yang/Lei/2015a}.

%
%
\textbf{Vision-based approaches} exploit light-sensitive sensors for the vehicle classification. Therefore, they are impacted by the lighting conditions, which results in a time-of-day-related performance.
%
Camera-based systems are the most prominent sensors of this category due to their high detection and classification accuracy, which can even be used for identifying vehicle make and models \cite{Siddiqui/etal/2015a}. However, the deployment of those systems might be complicated due to the involved regulations about potentially privacy-violating image data. Depending on the deployment type, vision-based approaches might be sensitive to congested traffic as individual vehicles might be occluded by other traffic participants. Image-based classification is a classic application field of \ac{ANN}-based methods and deep learning \cite{Liang/etal/2015a}. Recently, this approach has explicitly been subject to a new form of fraud attempts --\emph{adversarial examples} \cite{Eykholt/etal/2018a} -- where maliciously crafted inputs cause the neural networks to make incorrect predictions. The authors illustrate that adversarial attacks can change the prediction from, e.g., a stop sign to a fruit. It is important to understand that this example is deliberately chosen by the authors to emphasize that adversarial attacks do not require that the true class and the (mistakenly) predicted class are related to each other in any comprehensible way. In the real world, such an attack can be much more subtle, e.g., manipulating the sensors such that different vehicle classes might be confused, for example aiming to avoid payments in automatic toll collection services.
%
%
In \cite{Dong/etal/2015a}, a semi-supervised \ac{CNN} achieves 95.7\% accuracy for six vehicle classes on a daylight data set and 88.8\% on a nightlight data set.
%
%
For achieving a better computational efficiency, camera-based systems are often combined with other traffic sensors, which are then applied for performing the vehicle detection task. EasiSee is an example system proposed by Wang et al. \cite{Wang/etal/2014a}, which relies on \ac{AMR}-based detection.
%
%
Recently, aerial imaging \cite{Liu/Mattyus/2015a} based on small-scale \acp{UAV} has emerged as a novel traffic sensing approach. The integration of these vehicles into future \ac{ITS} is highly being discussed \cite{Menouar/etal/2017a} and might catalyze a novel development for dynamic sensing systems.
Apart from camera systems, other less popular approaches rely on \ac{PIR} \cite{Odat/etal/2018a}, \ac{LIDAR} \cite{Lee/Coifman/2015a} and laser scanning \cite{Xiao/etal/2016a}.

%
%
\begin{figure}[b]
	\centering
	\centering
	\includegraphics[width=1\columnwidth]{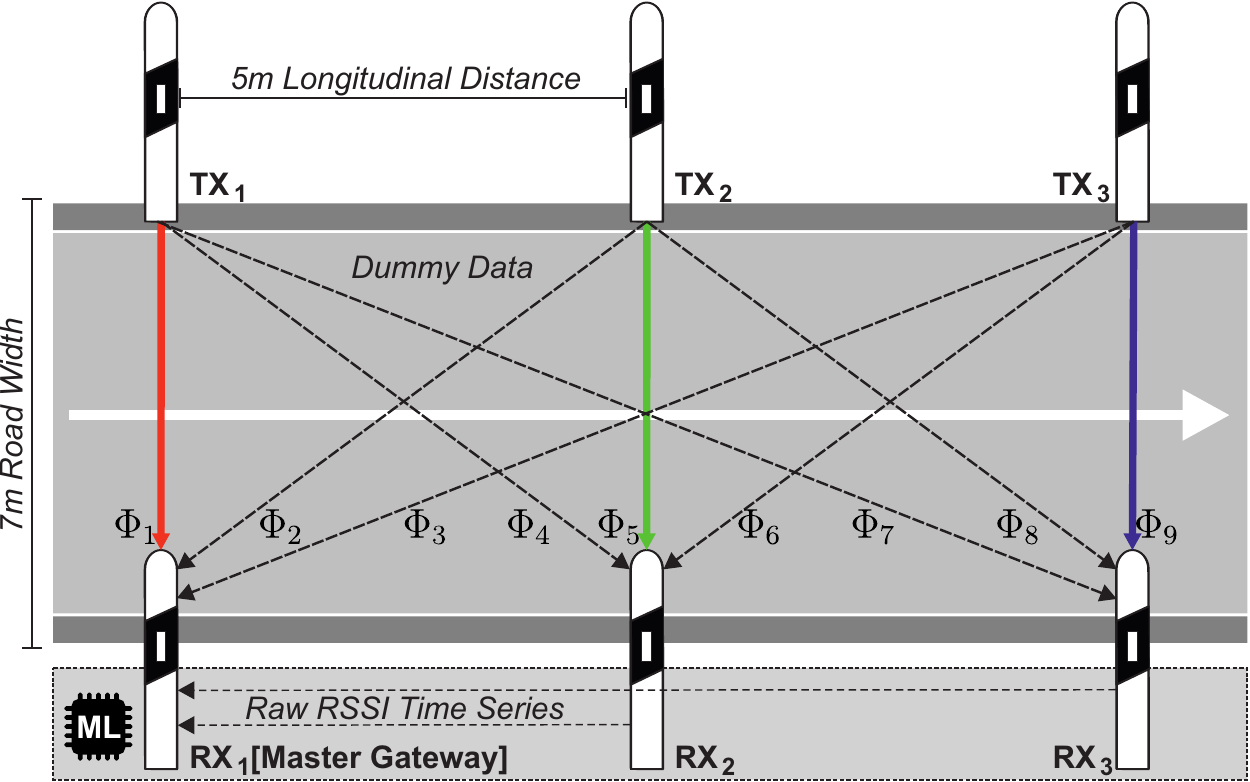}
	\caption{Schematic system overview. Each delineator post contains a communicating wireless sensor node. In total, nine different radio links $\Phi_{i}$ are present in the system.}
	\label{fig:schematics}
\end{figure}
%
%
\begin{figure}[b]
	\centering
	\includegraphics[width=1\columnwidth]{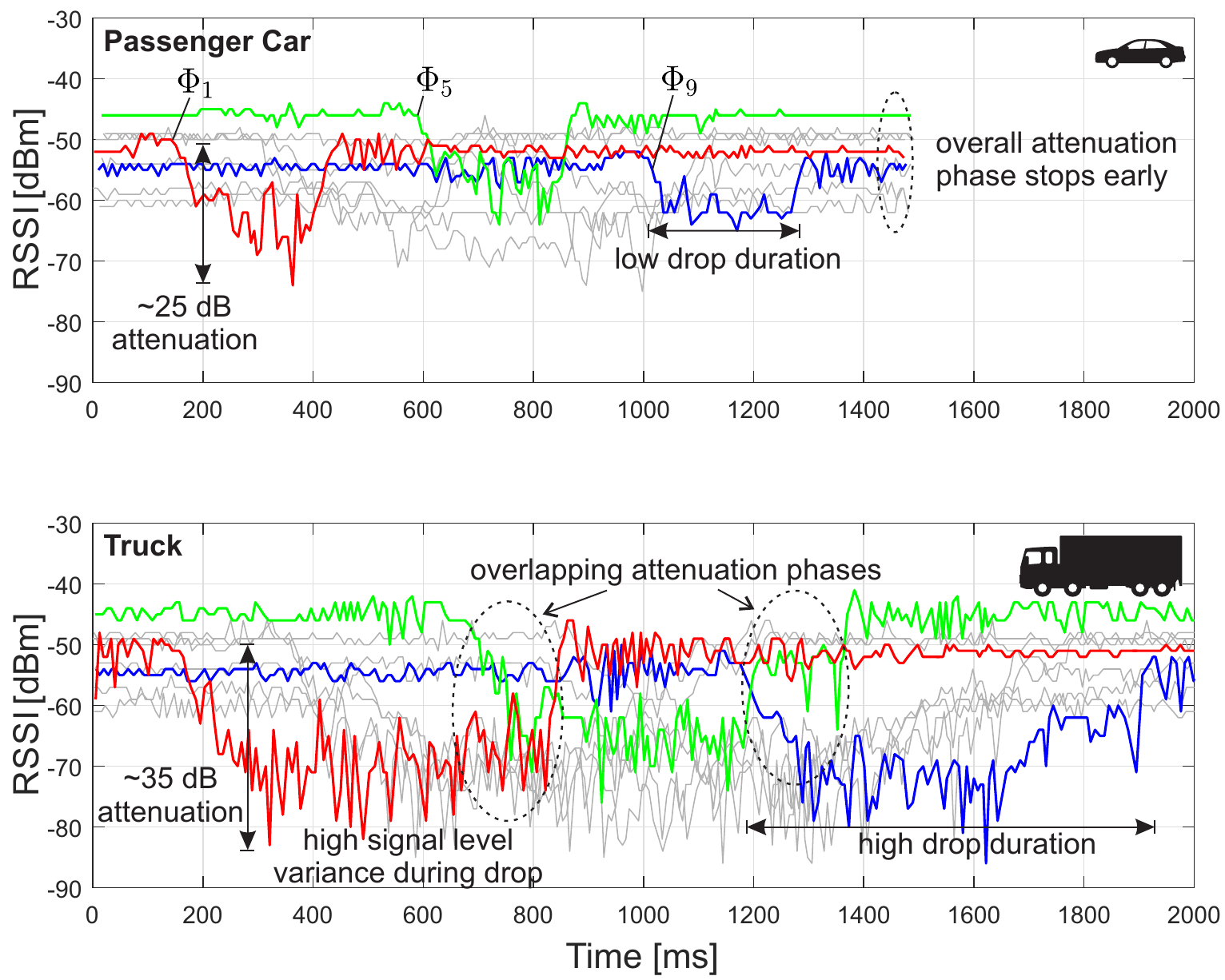}
	\caption{Examples for the type-specific multidimensional radio fingerprints which are related to the attenuation patterns of a vehicle passing the system installation. The highlighted signals are related to the direct links, gray signals correspond to diagonal links.}
	\label{fig:fingerprints}
\end{figure}

%
%
\textbf{Radio-based approaches} exploit the general ideas of \emph{radio tomography} and \ac{RTI} \cite{Adib/Katabi/2013a} for performing the classification task. Different communicating nodes are configured as a \ac{WSN} \cite{Anderson/etal/2014a}. When a vehicle passes the system, it acts as an obstacles, which causes an attenuation of the radio signal. The attenuation pattern -- which will be further referred to as the \emph{radio-fingerprint} -- is related to the vehicle's shape and material and can therefore be exploited for distinguishing different vehicle classes.

%
%
Existing systems rely on different communication technologies, which mostly operate in the 2.4~GHz band, e.g., Bluetooth \cite{Friesen/McLeod/2015a, Bernas/etal/2018a} or \ac{CSI}-based WiFi \cite{Won/etal/2018a}. They are robust against challenging weather conditions as the signal propagation in the considered frequency band is hardly impacted by rain and snowfall \cite{Crane/1980a, Rappaport/etal/2019a}. However, the performance of those systems relies on the presence of a  sufficient \ac{SNR} which may require environment-specific calibrations of the transmission parameters . In addition, interference might impact the expected behavior if the systems are deployed in the unlicensed frequency spectrum with dense usage. In addition, radio jammers could be used to explicitly attack those systems.
Sardar et al. \cite{Sardar/etal/2019a} furthermore present an unique approach, which implements the general idea of radio-fingerprinting for long-range communication via \ac{LTE}. Their proposed  \ac{SDR}-based method entitled \emph{LTE-CommSense} monitors the received downlink \ac{CSI} the \ac{UE} side. The system is able to distinguish between three different vehicles with an average accuracy of 92.6\%. 

%
%
In addition to \ac{WSN}-based approaches, \ac{RADAR} systems \cite{Urazghildiiev/etal/2007a} are also applied for vehicle classification. In contrast to attenuation-based approaches, \ac{RADAR} systems operate on reflections of radio signals.

\section{Radio Fingerprinting-based Vehicle Detection and Classification} \label{sec:approach}

In this section, we present the system model and the individual components of the proposed radio-based single lane vehicle detection and classification system. 
%
%
The latter consists of six communicating sensor nodes, whereas three nodes are configured as \emph{transmitters} and are installed on the opposite road side of the three \emph{receiver} nodes. A schematic overview of the system configuration is shown in Fig.~\ref{fig:schematics}. The longitudinal spacing is set to a constant distance $\Delta d_{\text{lon}}$.

%
%
All nodes are based on low-cost off-the-shelf embedded computers, which are equipped with low power IEEE 802.15.4 radio modules. The latter operate in the 2.4~GHz band, use omnidirectional antennas, and apply a transmission power of 2.5~dBm. All transmitting nodes periodically send dummy data, which is exploited by the receivers to determine the \ac{RSSI} values of all radio links $\Phi_{i}$ with a sampling time of 8~ms. The channel access is performed based on a token ring mechanism.
%
%
If a vehicle is present in the area covered by the proposed system, it acts like an obstacle and causes attenuations for one or more of the radio signals. The information about all signals is aggregated at the \emph{master gateway} node, which provides the time-base for the measurements and performs the online application of the trained machine learning models. The multi-dimensional attenuation pattern forms the \emph{radio fingerprint}, which is the foundation for the vehicle classification, as it is highly depending on the shape and the material of the vehicle. Fig.~\ref{fig:fingerprints} shows example radio fingerprints for a passenger car and a truck.

\begin{figure*}[]
	\centering		  
	\includegraphics[width=1\textwidth]{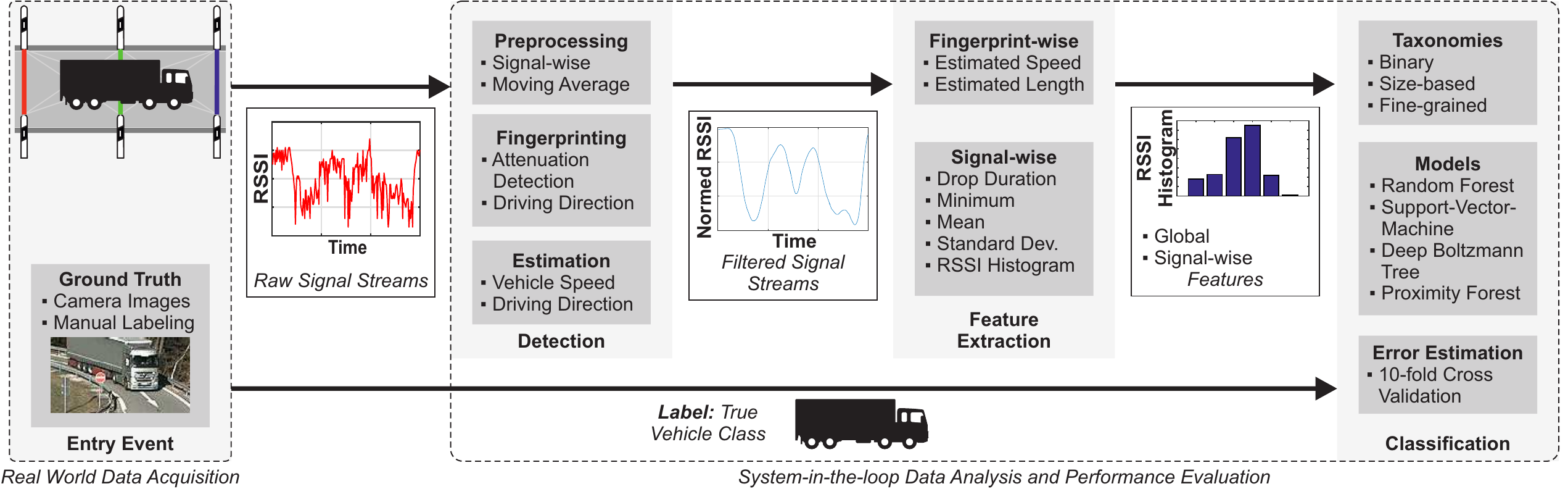}
	\caption{System architecture model and data procession pipeline for the machine learning-based vehicle classification using radio fingerprint information.}
	\label{fig:approach}
\end{figure*}

%
%
The different stages of the classification process are illustrated in Fig.~\ref{fig:approach}. During the real world data acquisition phase, time series data of all nine signals $\Phi_{i}$ is captured. The feature extraction is performed signal-wise on the filtered signals.
%
%
In addition, and only during the initial training phase, video images are captured as ground truth in order to allow the manual labeling of the captured radio fingerprints with the observed true vehicle class. 
%
%
The classification accuracy is then determined for different taxonomies and machine learning models.

%
%

\subsection{Preprocessing and Vehicle Detection}

Before the machine learning-based classification can be applied, the presence of a vehicle needs to be detected and the actual radio fingerprinting information needs to be extracted from the data streams.
%
%
In an initial step, all signals are normalized to their idle level $\hat{\Phi}_{i}$ and filtered with a moving average filter of size $N$
\begin{equation}
	\bar{\Phi}_{i}(k) = \frac{1}{N} \sum_{j=0}^{N-1} \frac{\Phi_{i}(k-j)}{\hat{\Phi}_{i}}
\end{equation}
in order to reduce the effect of short-term fluctuations and compensate potential power offsets.

%
%
Afterwards, the start $\attStart{i}$ and the end $\attEnd{i}$ of the attenuation phase are detected based on a threshold mechanism. The transition decision between the states \texttt{Attenuated} and \texttt{Not Attenuated} is delayed by a guard interval of $w$ steps, in order to allow the compensation of local anomalies which can only be detected after sufficient data has been received. 
%
%
\fig{b}{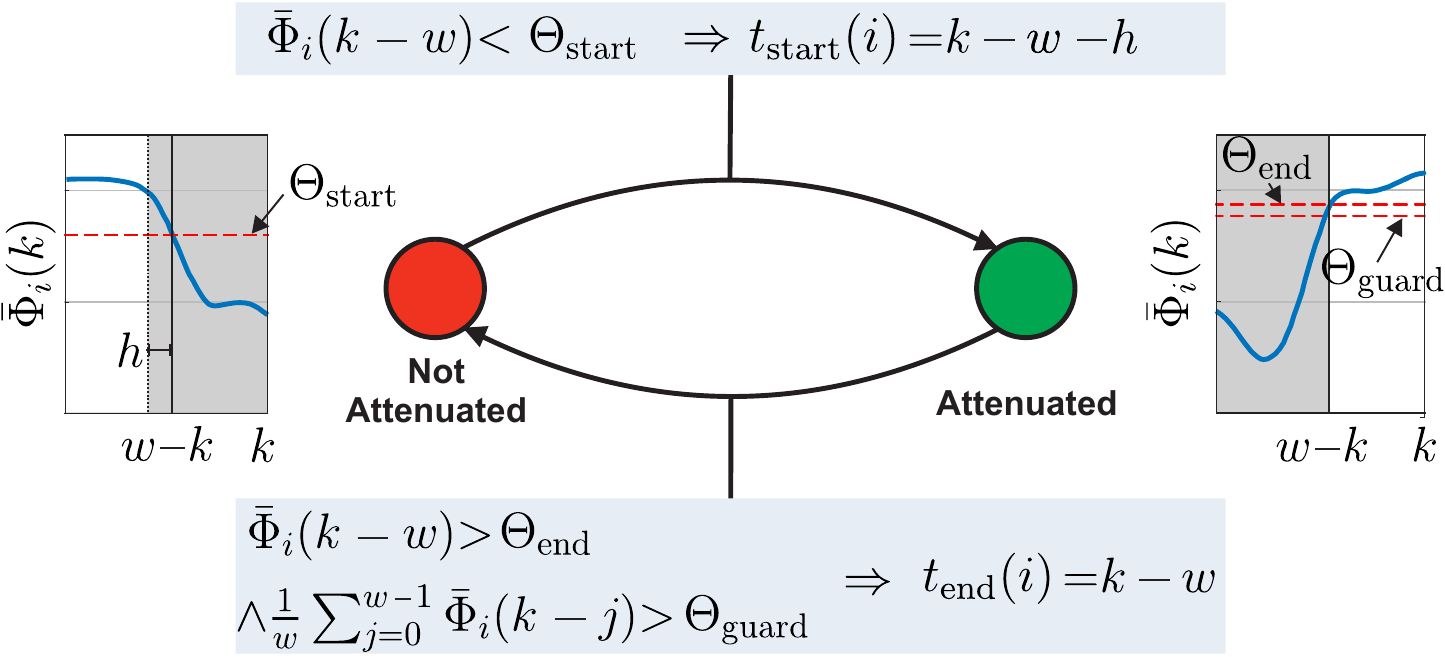}{State machine, transition conditions and triggered actions for the signal-wise attenuation detection.}{fig:detector}
The resulting state machine of the detector is illustrated in Fig.~\ref{fig:detector}.
%
%
The attenuation start $\attStart{i}$ is detected if the signal at $k-w$ undercuts the threshold $\thStart$. However, since this point marks an already active attenuation phase, $h$ previous data points are added to the signal in order to also consider the actual turning point.
%
%
The detection of the attenuation end $\attEnd{i}$ is performed with respect to the threshold $\thEnd$. As a second condition, the average signal level within the guard interval $w$ needs to be at least $\thGuard$ in order to make a definite decision. This is important as the radio fingerprints of different vehicle types (e.g., truck with trailer) might consist of multiple attenuation phases. 
%
%
The association between the attenuation phases and the vehicles within the area covered by the system is performed based on attenuation order of the different links. After the detector has transited from \texttt{Attenuated} to \texttt{Not Attenuated}, it updates the internally used abstract vehicle identifier.

%
%
In addition to detecting the presence of a vehicle in the covered area, the vehicle's driving direction can be estimated for allowing the detection \emph{wrong-way} drivers. Using knowledge about the distances between the node deployments, the average velocity $\tilde{v}$ of the vehicle can be estimated considering the times when the vehicle passes the straight links $\Phi_{1}$, $\Phi_{5}$ and $\Phi_{9}$
%
%
\begin{align}
	\tilde{v} = \frac{1}{3} \left( 
	\frac{d(1,5)}{\Delta t(1,5)} + 
	\frac{d(1,9)}{\Delta t(1,9)} + 
	\frac{d(5,9)}{\Delta t(5,9)} 
	\right) 
\end{align}
with $\Delta{t}(i,j) = \attStart{j} - \attStart{i}$ and $d(i,j)$ being the longitudinal spacing between the transmitter nodes of the radio signals $\Phi_{i}$ and $\Phi_{j}$. As wrong-way drivers enter the system in the opposite direction, the attenuation order of the different links is reversed and $\tilde{v}<0$.

%
%
\subsection{Feature Extraction}
The overall considered feature set consists of 92 different indicators. It includes 10 features for each of the 9 radio links as well as the globally derived estimations for the vehicle speed and length. The latter is computed based on knowledge about the system dimensions 
%
%
\begin{align}
	\tilde{l} = \frac{|\tilde{v}|}{3} \left( 
	\durAtt{1} +
	\durAtt{5} +
	\durAtt{9}
	\right) 
\end{align}
with $\durAtt{i} = \attEnd{i}-\attStart{i}$ being the duration of the attenuation phase for signal $i$.
%
%
The per-signal part of the feature vector consists of purely mathematical features: Duration, minimum, mean, and standard deviation of the filtered signal $\bar{\Phi}$ during the attenuation phase. Tab.~\ref{tab:feature_statistics} provides an overview about the class-specific feature values for the binary classification task for $\Phi_{1}$. Especially for the speed, length, attenuation duration and minimum signal value, significant differences between the two vehicle types can be observed.

\newcolumntype{z}{>{\centering\arraybackslash}m{0.25in}}
\newcolumntype{Z}{>{\centering\arraybackslash}m{0.33in}}
\begin{table}[ht]
	\centering
	\caption{Statistical Properties of the Features Set for $\Phi_{1}$ for Binary Vehicle Classification }
	\begin{tabular}{zZZZZZZ}
		
		\toprule
		\textbf{Class} & $\mathbf{\tilde{v}}$ \textbf{[km/h]}  & $\mathbf{\tilde{l}}$ \textbf{[m]} & $\mathbf{\durAtt{1}}$ \textbf{[s]} & $\mathbf{min}$ $(\bar{\Phi}_{1})$ & $\mathbf{mean}$ $(\bar{\Phi}_{1})$ & $\mathbf{std}$ $(\bar{\Phi}_{1})$ \\
		
		\midrule
		\multirow{2}{0.5cm}{\textbf{Car-like}}
		& $40.47$ & $5.22$ & $0.46$ & $0.72$ & $0.86$ & $0.012$ \\
		& $\pm7.2$ & $\pm1.08$ & $\pm0.11$ & $\pm0.06$ & $\pm0.03$ & $\pm0.005$ \\
		
		\multirow{2}{0.5cm}{\textbf{Truck-like}}
		& $31.42$ & $16.53$ & $1.9$ & $0.62$ & $0.77$ & $0.01$ \\
		& $\pm5.4$ & $\pm3.3$ & $\pm0.5$ & $\pm0.05$ & $\pm0.03$ & $\pm0.003$ \\		
		
		\bottomrule
	\end{tabular}
	\label{tab:feature_statistics}
\end{table}

In addition, a normalized histogram of the \ac{RSSI} values is included, which consists of 6 bins.
%
%
\fig{}{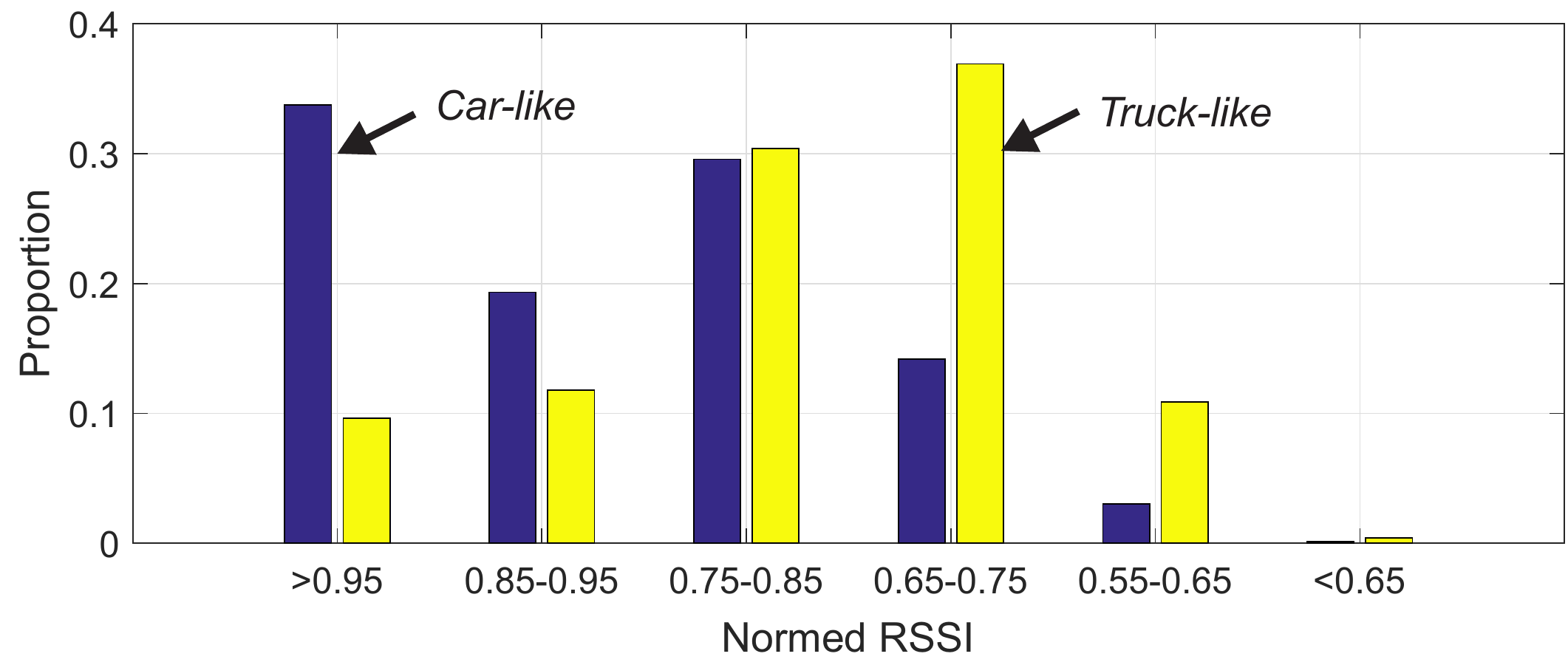}{Overall normed \ac{RSSI} histograms for car-like and truck-like vehicles, which are contained in the signal-specific feature set.}{fig:histograms}
Fig.~\ref{fig:histograms} shows the overall histograms of the car-like and truck-like classes with respect to the binary classification scheme.

\subsection{Classification Taxonomies}\label{sec:taxonomies}
%
%
\fig{}{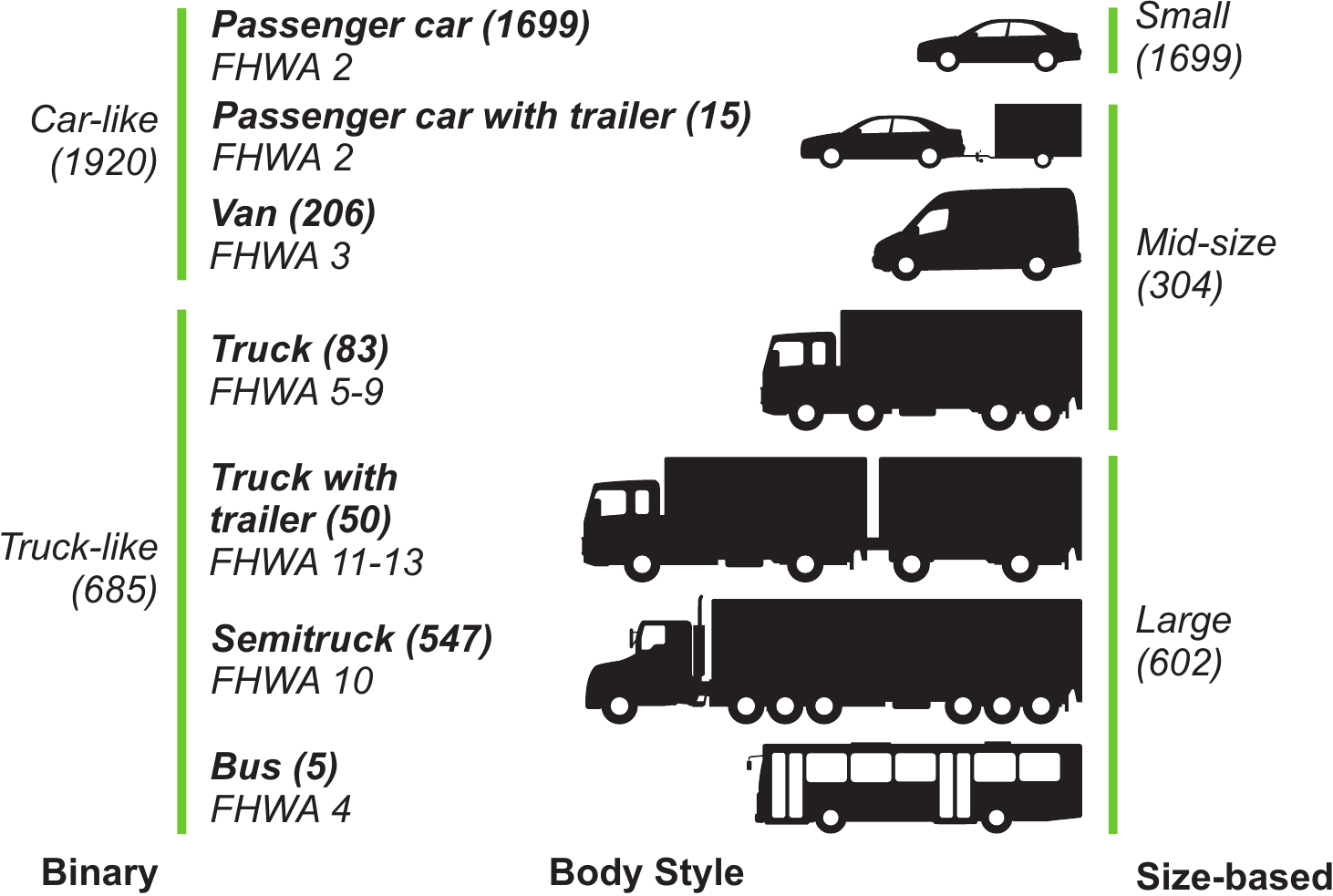}{Overview about the different vehicles classes and number of measurements for the considered three taxonomies.}{fig:taxonomy}
In this paper, we consider multiple taxonomies, for which the covered vehicle classes are illustrated in Fig.~\ref{fig:taxonomy}. The considered definitions are chosen application-specific (e.g., distinguishing cars and trucks is important for length-based parking space accounting). Since the radio-based classification approach relies on extracting information about the general vehicle \emph{shape} from attenuation patterns of radio signals, it is not designed for working well with axle count-based classification taxonomies (e.g., the 13-category \ac{FHWA} scheme \cite{Klein/etal/2006a}). Hereby, vehicles of similar shape might be assigned to different classes because of their axle count. In the following, we consider the following taxonomies:
\begin{itemize}
	\item \textbf{Binary} with two subclasses \emph{car-like} and \emph{truck-like}.
	\item \textbf{Size-based}with three different subclasses \emph{small}, \emph{mid-size} and \emph{large}.
	\item \textbf{Body style} is related the vehicle's shape and consists of seven different subclasses.
\end{itemize}
For the body style taxonomy it is remarked that the shapes of the different vehicle classes are subject to considerable overlap, e.g., buses and semitrucks. 
Due to this distortion between vehicle shape and corresponding class, it should be expected that all machine learning methods are subject to a certain non-reducible amount of class confusion. 
Nevertheless, the body style taxonomy is included in order to get a better understanding of the limitations of the proposed system. 
For the bus class in the body style taxonomy, it is remarked that the number of measurement samples is below the number of cross validation folds. Due to this underrepresentation of the bus class, the estimation of the classification accuracy is inherently pessimistic.

\subsection{Classification Models}

For the machine learning-based vehicle classification, we compare different established and state-of-the-art machine learning models:
%
%

%
%
\textbf{\acfp{RF}}
\cite{Breiman/2001a} are constructed from a set of bootstrapped \cite{Efron/1979a} \ac{CART} trees. Each tree is learned from a random sub-set of the training data, using only a random sub-set of all features. For inference, the predictions of all trees are aggregated. E.g., in case of classification, the random forest predicts the majority class that is predicted by most of the trees. The model is parameterized to analyze 100 random trees.
%
%

\textbf{\acfp{SVM}} \cite{Cortes/Vapnik/1995a} learn a hyperplane that separates real-valued data points in a $d$-dimensional space ${\rm I\!R}^d$ by minimizing a specific objective function. The hyperplane cuts ${\rm I\!R}^d$ into half spaces, whereas each side of the hyperplane contains the majority of data points that belong to one of two classes. In order to apply \acp{SVM} to more than two classes, the one-vs-all strategy is employed. \ac{SVM}-based classification is achieved with $f(x)=\text{sign}(\langle\mathbf{\beta}, \mathbf{x}\rangle)$ with a weight vector $\beta \in {\rm I\!R}^d$.
As discussed in Sec.~\ref{sec:related_work}, \acp{SVM} are widely used in established classification systems.
In a pre-evaluation step, we evaluated different model variants (e.g., \acp{SVM} with \ac{RBF} and polynomial kernels), whereas the classic linear $l_{2}$-regularized \ac{SVM} achieved the highest average accuracy. Its objective function is written as
\begin{equation}
\min_\mathcal{\beta} \frac{1}{2} || \mathcal{\beta}||^{2}_{2} + C \sum_{\left(y,\mathbf{x} \right) \in \mathcal{D}} \underbrace{\max \left\lbrace 0,1-y \left\langle \mathcal{\beta}, \mathbf{x} \right\rangle  \right\rbrace }_{\text{Hinge Loss}}
\end{equation}
with $C$ being a parameter for controlling the trade-off between correctly classified training samples and the model complexity.

%
%
\textbf{\acfp{DBT}}
belong to the class of deep learning models \cite{Goodfellow/etal/2016a}. While classic (non-deep) approaches rely on hand-crafted features or user defined feature transformations (e.g., the \ac{RBF} kernel), deep learning methods aim at phrasing almost all parts of the model as differentiable functions. Thus, numerical optimization methods can replace what was formerly done by hand. However, these methods have at least two significant drawbacks: First, the user has to choose various hyper-parameters, including learning rates, activation functions, and the network architecture itself. Selecting good hyper-parameters is tedious and time-consuming and may lead to highly overfitted models if not done carefully. Second, deep models contain many learnable weights and hence require a large amount of data. While the second issue is a statistical necessity, the problem of choosing good hyper-parameters can be solved by resorting to theoretically sound statistical models. A good compromise are \acp{DBM} \cite{Salakhutdinov/Hinton/2009a}---they constitute the cut-set between deep models and statistical sound models. Learning these models exactly is computationally intractable in general, but recent advances in machine learning \cite{Piatkowski/2019a} allow for an efficient learning of a sub-class of \acp{DBM}, the \acp{DBT}. Here, we use \acp{DBT} to estimate the joint probability mass function $\mathbb{P}(\boldsymbol{X},Y)$ of features $\boldsymbol{X}$ and class label $Y$, and apply Bayes' rule to predict the most likely class $\arg\max_i\mathbb{P}(y_i \mid \boldsymbol{X})$. The generative nature of the model further allows for solving additional tasks like data synthesis and the detection of sensor faults.

%
%
\textbf{\acfp{PF}} define the new state-of-the-art in distance-based time series classification \cite{Lucas/etal/2019a}. Like \acp{RF}, \acp{PF} are a set of tree classifiers but instead of \ac{CART} trees, proximity trees are used. Those differ from ordinary decision trees in the tests applied at internal
nodes. Whereas a regular decision tree applies a test based on the value of
an attribute (e.g. if $\text{\ac{RSSI}} < -102~dB$, follow the left branch, otherwise follow
the right branch), each branch of an internal node of a proximity tree has
an associated data point from the training set and an object follows the branch corresponding to
the exemplar to which it is closest according to a parameterized similarity
measure. See \cite{Lucas/etal/2019a} for details on how internal data points and measures are chosen. 
While the complexity of the former leading method is bounded by $\mathcal{O}(l^4 n^2)$ for $n$ time-series of length $l$, \acp{PF} have an average runtime complexity of $\mathcal{O}(l^2 n \log n )$. Especially the reduction of the depdence on the length $l$ by two orders of magnitude enables the use of \acp{PF} in various practical settings. Recall that for our \ac{RSSI} series, we have $l=400$---a length that renders any methods which scale as $l^4$ intractable.

\section{Evaluation Methodology} \label{sec:methods}

In this section, the methodological setup for the real world data acquisition and the quality measures for the machine learning-based vehicle classification are introduced.

\subsection{Real World Data Acquisition}

\begin{figure}[b] 
	\centering
	\includegraphics[height=3.5cm]{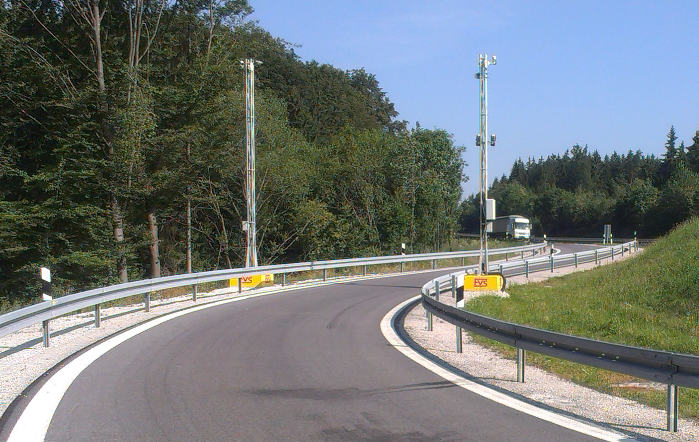}
	\includegraphics[height=3.5cm]{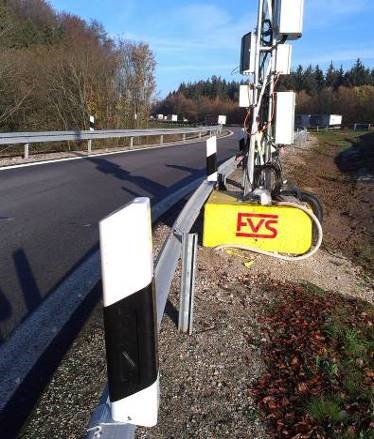}
	\caption{Experimental live deployment of the system on the German highway A9 within an official test field by the German Federal Ministry of Transport and Digital Infrastructure.}
	\label{fig:field_deployment}	
\end{figure}

For the real world data acquisition, a live system has been deployed within an official test field of the German Federal Ministry of Transport and Digital Infrastructure at a rest area at the German highway A9. The real world deployment of the system and the sensor nodes is shown in Fig.~\ref{fig:field_deployment}. The \ac{RSSI} values are represented as continuous data streams. In total, measurement data for 2605 vehicles was captured and labeled manually based on camera images.
%
%
A summary about the parameter definitions is shown in Tab.~\ref{tab:parameters}.
%
%
\newcommand{\entry}[2]{#1 & #2 \\}
\newcommand{\head}[2]{\toprule \entry{\textbf{#1}}{\textbf{#2}}\midrule}

\begin{table}[ht]
	\centering
	\caption{Definition of the System Parameters}
	\begin{tabularx}{\columnwidth}{Xr}
		
		\head{Parameter}{Value}
		
		\entry{Longitudinal node spacing $\Delta d_{\text{lon}}$}{5~m}
		\entry{Moving average filter size $N$}{10}
		\entry{Transition decision guard interval $w$}{10}
		\entry{Attenuation start offset $h$}{5}
		\entry{Attenuation start threshold $\thStart$}{0.92}
		\entry{Attenuation end threshold $\thEnd$}{0.975}
		\entry{Attenuation guard interval threshold $\thGuard$}{0.95}
		\bottomrule
		
	\end{tabularx}
	\label{tab:parameters}
\end{table}

\subsection{System-in-the-loop Simulation} \label{sec:simulation}

The system evaluation is performed based on a system-in-the-loop simulation setup, where the acquired raw continuous \ac{RSSI} data streams are fed into the processing pipeline for detection and classification according to Fig.~\ref{fig:approach}. 
This way, new detection and classification models can be developed in a flexible way based on the reliable foundation of the acquired real world data. Furthermore, it allows to virtually manipulate system properties in postprocessing evaluations (e.g., for analyzing the detection of the wrong-way drivers in Sec.~\ref{sec:results_detection}).

\subsection{Quality Measures for Vehicle Classification}

For assessing the quality of a machine learning model $f$ on a data set $\mathcal{D}$, we calculate the \emph{accuracy} (which is also referred to as \emph{classification success ratio} in related work) as
%
%
\begin{equation}
\text{ACC}(f;\mathcal{D}) = \frac{1}{|\mathcal{D}|} \sum_{(y, \mathbf{x})\in\mathcal{D}}1_{\left\lbrace y=f(\mathbf{x})\right\rbrace}
\end{equation}
with $|\mathcal{D}|$ being the cardinality of $\mathcal{D}$ and $1_{\left\lbrace y=f(\mathbf{x})\right\rbrace}$ being the indicator function that only evaluates to 1 if $f(\mathbf{x})$ outputs the correct class $y$ and is 0 otherwise.
When $|\mathcal{D}|$ approaches $\infty$, $\text{ACC}(f;\mathcal{D})$ converges to the probability of doing a correct classification: $\lim_{|\mathcal{D}|\rightarrow\infty} \text{ACC}(f;\mathcal{D}) = {\rm I\!P}(f(\mathbf{x})=Y)$.
However, $\text{ACC}(f;\mathcal{D})$ will only be a reliable estimate for ${\rm I\!P}(f(\mathbf{x})=Y)$ when $\mathcal{D}$ consists of data, which was not used to learn $f$.
Therefore, $\mathcal{D}$ is split into a training set $\mathcal{D}_{\text{train}}$, which is used to learn the model $f$, and a test set $\mathcal{D}_{\text{test}}$, which is used to assess the quality of $f$. Still, a particular choice of $\mathcal{D}_{\text{train}}$ and $\mathcal{D}_{\text{test}}$ might result in overly pessimistic or overly confident accuracy estimates.
%
%
To overcome this issue, $\mathcal{D}$ is partitioned into $k$ subsets $\mathcal{D}_{1}, ..., \mathcal{D}_{k}$. The learning process is performed $k$ times, whereas in each run $\mathcal{D}_{i}$ is utilized as the test set $\mathcal{D}_{\text{test}}$ to compute $\text{ACC}(f;\mathcal{D}_{i})$ and the remaining subsets jointly form the training set $\mathcal{D}_{\text{train}}$. The $k$-fold cross validated accuracy is then computed as the average of the $k$ folds with
%
%
\begin{equation}
\text{ACC}_{k} = \frac{1}{k} \sum_{i=1}^{k} \text{ACC}(f;\mathcal{D}_{i})
\end{equation}
which provides a more reliable measurement for assessing the performance on unknown data. In addition, is allows to calculate the standard deviation of the $k$ validation runs with 
%
%
\begin{equation}
\hat{\sigma}_{\text{ACC}} = \sqrt{\left( \frac{1}{k} \sum_{i=1}^{k}\text{ACC}(f; \mathcal{D}_{i})^{2} \right) - \text{ACC}^{2}_{k}}
\end{equation}
which quantifies the uncertainties of the accuracy estimation. In this paper, we apply $k=10$ in the following evaluations.

\subsection{Feature Importance Analysis for the SVM Classifier}

Since the \ac{SVM} classifier turned out to achieve the highest classification accuracy in most settings, Sec.~\ref{sec:results_classification} focuses on a deeper investigation of this classifier. In order to allow us to analyze the relative radio link importance for the overall classification result in the result section, we first define a novel system- and \ac{SVM}-specific method for the relative feature importance. 

Recall that the SVM decision is computed via $f(x)=\text{sign}(\langle\mathbf{\beta}, \mathbf{x}\rangle)$. 
Here, $\mathbf{x}$ contains the group of global features (G), including vehicle speed and length, as well as a group of 10 features per radio link $\Phi_j$. 
We denote the set of feature groups by $\mathbb{G}=\{ \text{G}, \Phi_1, \Phi_2, \dots, \Phi_9\}$. 
Each feature weight $\mathbf{\beta}_i$ can be decomposed into two components: (1) its sign $\text{sign}(\mathbf{\beta}_i)$ which points into the 
direction of one of the two half spaces, and (2) its magnitude $|\mathbf{\beta}_i|$ which indicates how strong it points into that direction. 
Both components are now exploited to compute a new measure for SVM feature importance. For the best of our knowledge, this measure never appeared in the literature before. 
Let $g : \{1,2,\dots,d\} \to \mathbb{G}$ indicate to which feature group the $i$-th feature belongs. 

In case of binary classification, the importance of group $j \in \mathbb{G}$ on class $y\in\{-1,+1\}$ is computed via
\begin{equation*}
I(j,y) = \frac{1}{Z} \sum_{i : g(i) = j} |\mathbf{\beta}_{i}| \times 1_{\{ y = \text{sign}(\mathbf{\beta}_i) \}}
\end{equation*}
with $Z=\sum_{i : g(i) = j} |\mathbf{\beta}_{i}|$. $I$ is normalized, w.r.t. $y$, such that $I(j,+1) + I(j,-1) = 1$. 

In case of $Y>2$ classes, $K={Y(Y-1)}/{2}$ SVMs are computed for using pairwise classification (aka one-vs-one). 
Thus, let $\mathbf{\beta}_{i,k}$ denote the $i$-th feature weight of the $k$-th SVM. 
Moreover, a mapping $\gamma$ is required which transfers an SVM index and a binary class index to the actual multi-class index, 
i.e., $\gamma : \{1,2,\dots,K\}\times\{-1,+1\}\to\{1,2,\dots,Y\}$.  
The feature importance is then computed via
\begin{equation*}
I(j,y) = \frac{1}{Z} \sum_{i : g(i) = j} \sum_{k=1}^{K} |\mathbf{\beta}_{i,k}| \times 1_{\{ y = \gamma(j,\text{sign}(\mathbf{\beta}_{i,k})) \}}
\end{equation*}
with $Z=\sum_{i : g(i) = j} \sum_{k=1}^{K}|\mathbf{\beta}_{i,k}|$. Again, $I$ is normalized, i.e., $\sum_{y=1}^Y I(j,y) = 1$. 

The higher the score $I(j,y)$, the more important is feature group $j$ for predicting class $y$. However, care has to be taken when these scores are interpreted: 
In order to get meaningful results, the SVM model has to be trained from {\em normalized} data, such that $\mathbf{x}\in[-1;1]^d$.
If this requirement is violated, model weights for different features will have different scale which renders the accumulation of weights for different features invalid. 

\subsection{Machine Learning Tools}

%
%
For training \ac{SVM} and \ac{RF} models, the \ac{LIMITS} \cite{Sliwa/etal/2020c} framework is used which provides high-level data analysis automation based on validated \ac{WEKA} \cite{Hall/etal/2009a} model implementations.
\ac{DBT}s are learned via the \texttt{px} framework\footnote{\url{https://randomfields.org/px}}. 
%
%
Proximity forests are trained with the original implementation provided by the authors \cite{Lucas/etal/2019a}.
%
%
In order to eliminate the effects of randomization and to allow a fair comparison between results generated with different tools, the data subsets for the cross validation are generated externally and are shared by all machine learning tools.


\section{Results} \label{sec:results}

In this section, we present and discuss the results for vehicle detection and classification.

\subsection{Detection of Vehicle Presence and Driving Direction} \label{sec:results_detection}

For evaluating the detection performance of the proposed system, we replay the time series streams of all 2605 traces in the system-in-the-loop evaluation setup, which is described in Sec.~\ref{sec:simulation}. Furthermore, for giving a first feasibility showcase about wrong-way driver detection capabilities of the proposed system, we virtually invert the order of the sensor nodes, such that each vehicle first passes $\Phi_{9}$ when entering the system and $\Phi_{1}$ upon leaving the covered area.

%
%
\fig{}{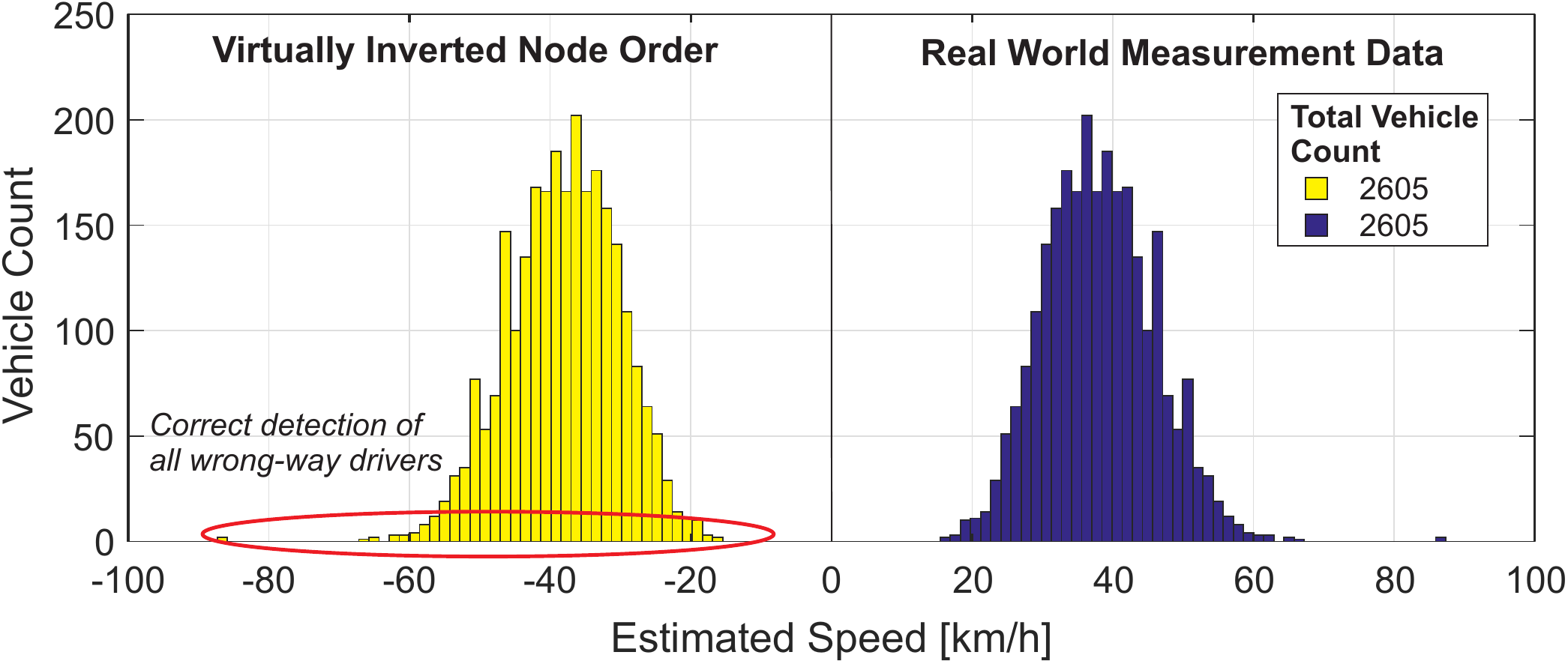}{Histograms of the speed estimations for the real world data and the artificially inverted node assignment. The vehicle count matches the ground truth and all wrong-way drivers are successfully detected.}{fig:speed}
%
%
Fig.~\ref{fig:speed} shows a histogram of the resulting speed estimations. Since the real world measurement setup does not capture a ground truth for the driving speed, the accuracy of the speed estimation cannot be assessed. However, it can be seen that the distributions for the two data sets are clearly divided and the driving direction is detected correctly for all vehicles by considering only the sign of the speed value. However, we remark the applied evaluation methodology only illustrates the general suitability of radio-based detection of wrong-way drivers. In order to make a quantified statement about the resulting detection accuracy of those special situation, a dedicated real world measurement campaign yet needs to be performed.

%
%
The resulting total vehicle count is equal to the number of vehicles traces, thus all vehicles are successfully detected. Even the detection accuracy per single signal $\Phi_{i}$ is 100\% in the vast majority of the evaluations. However, for $\Phi_{7}$, only  91\% of the passenger cars are detected correctly. $\Phi_{7}$ is one of the links with the highest signal attenuation due to the large distance between transmitter and receiver. In addition, the passenger cars represent the smallest vehicle type with the lowest attenuation.
However, these signal-specific outliers do not impact the overall detection capability of the proposed system due to the high grade of contained redundancies.

\subsection{Vehicle Classification} \label{sec:results_classification}

Based on the foundation of the detected vehicle fingerprints of Sec.~\ref{sec:results_detection}, the classification performance is analyzed based on 10-fold cross validation. Within each fold 90\% of the data is used for training and 10\% is used for testing. This mechanism is repeated ten times and the statistical deviations between the individual folds are analyzed.

The resulting impact of errors in the classification process is highly application-specific. While traffic management systems might be robust to misclassification as lane clearance decisions are based on relative vehicle densities, toll collection systems might require reliable distinction between trucks and passenger cars.
An overall comparison of the resulting classification accuracies for the different learning models and the considered taxonomies is shown in Fig.~\ref{fig:results_box}.
%
%
\fig{}{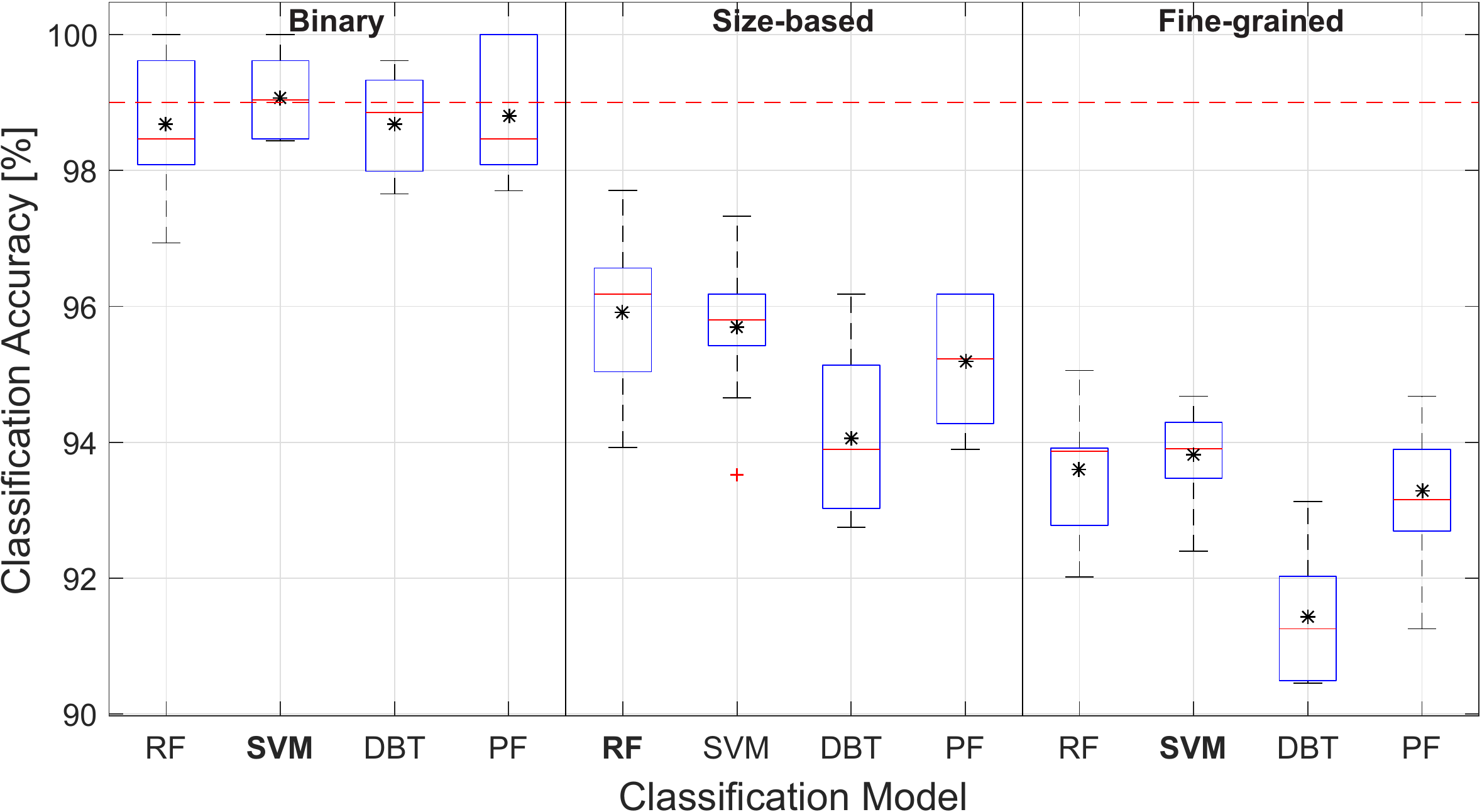}{Comparison of the overall classification accuracies for the considered machine learning models and vehicle classification schemes.}{fig:results_box}
The dashed line indicates a reference accuracy level of 99\% according to commonly used certification requirements for practical application scenarios where trucks and passenger cars need to be distinguished, e.g., for length-based parking space accounting. 
While all methods are capable of exceeding the given threshold in {\em some} runs in the {\em binary setting}, only the \ac{SVM} achieves this correct classification rate {\em on average}. 
Due to the stochastic nature of the \ac{RF} construction, cross validation standard errors exceed those of the deterministically trained methods. 
For the more complex taxonomies, the overall accuracy decreases for all models. 
With exception of the \ac{DBT} model, \ac{SVM}, \ac{RF}, and \ac{PF} achieve a similar level of performance. 
Note, however, that in contrast to the other methods, the \ac{DBT} is a full \emph{generative} model of the data for which the learning capabilities go beyond the scope of sole classification. That is, \acp{DBT} are able to estimate the probability measure that underlies the given data set. Based on this information and the derived statistical properties, new virtual data points can be generated, e.g., for being exploited in a future extension of the simulation setup. 

%
%
\basicFig{}{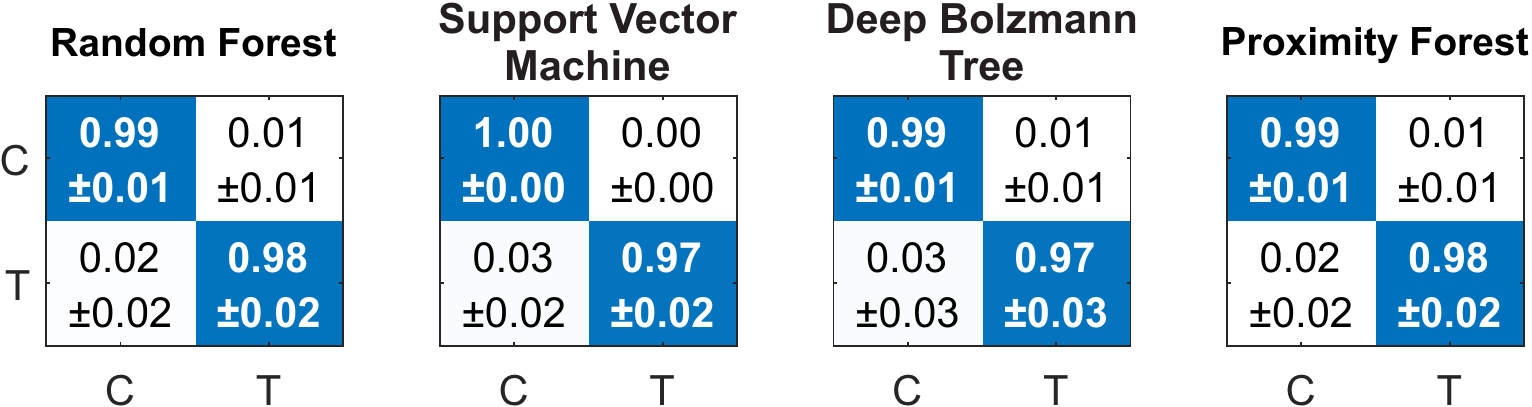}{Normalized confusion matrices for binary vehicle classification. \emph{C: Car-like, T: Truck-like}.}{fig:confusion_binary}{0cm}{0cm}{1}
%
%
%
%
%
\fig{}{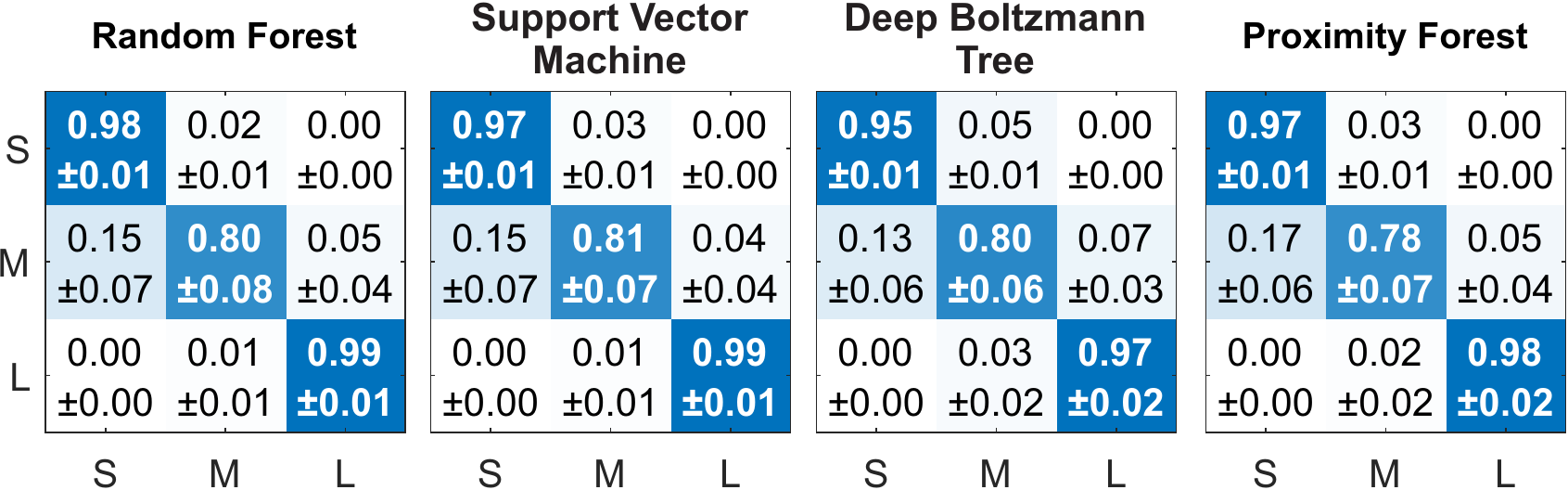}{Normalized confusion matrices for size-based vehicle classification. \emph{S: Small, M: Mid-size, L: Large}.}{fig:confusion_size}
%
%
%
%
%
\begin{figure*}[]
	\centering		  
	\includegraphics[width=1\textwidth]{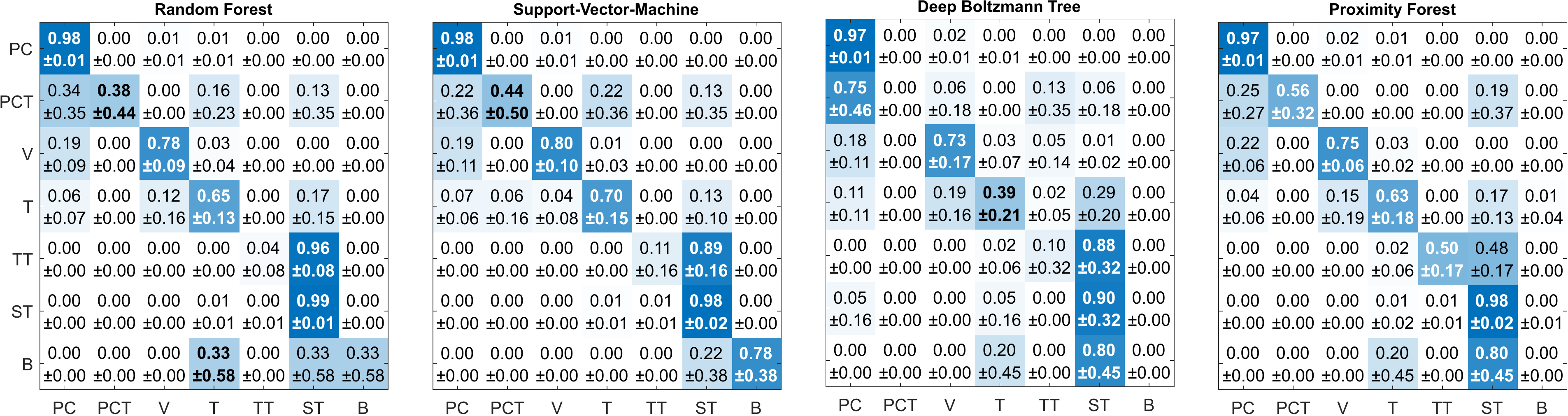}
	\caption{Normalized confusion matrices for body style vehicle classification. \emph{PC: Passenger car, PCT: Passenger car with trailer, V: Van, T: Truck, TT: Truck with trailer, ST: Semitruck, B: Bus.}}
	\label{fig:confusion_fine}
\end{figure*}

Fig.~\ref{fig:confusion_binary}, Fig.~\ref{fig:confusion_size}, and Fig.~\ref{fig:confusion_fine} show the class-specific performance for all three taxonomies and the considered machine learning models. 
In the binary case, for all models, the main part of the error arises from falsely classifying trucks as passenger cars. 
A closer look at the misclassified vehicles reveals that smaller trucks are confused with vans. 
As expected, the dominant source of failure for the size-based taxonomy manifests at the boundary of small and large vehicles, i.e., members of the mid-size class are falsely classified as small vehicles. 
For the more fine-grained taxonomies, the underrepresentation of some vehicle classes (e.g., bus) manifests in a larger standard deviation than for the better represent vehicles classes. In the body style taxonomy, the overall picture is not that clear: The results differ significantly among the machine learning methods.
Common misclassifications include the confusion between buses and semitrucks. As mentioned in Sec.~\ref{sec:taxonomies}, buses are underrepresented in the considered data set an not contained in all folds of the cross validation. Moreover, regular vehicle types are often confused with their trailered counterparts. 
As pointed out in Sec.~\ref{sec:taxonomies}, this phenomenon is implied by the overlap among some real world vehicle shapes.
However, in contrast to the feature-based approaches, the time-series-based proximity forest is least amenable to this type of failure. 
Recalling the definition of the considered taxonomies, it can be observed that classification errors rarely violate the boundaries of the their corresponding binary class.

\subsection{Feature Importance}

To get more insights about the classification performance, it is important to understand which radio link has the highest impact on the actual class prediction. Hence, we analyzed the model weights of the best performing model, the SVM. 
Fig.~\ref{fig:feature_importance} contains the importance scores for all 10 feature groups and all three taxonomies. 
In what follows, we focus on discussing the feature groups with the major impact on the classification performance of the respective taxonomies. 
In the binary and size-based taxonomies, where vehicle shapes are rather well-separated, the group of global features which contain estimates for vehicle length and speed, is of outstanding importance. 
In contrast, the SVMs for the body style taxonomy mainly rely on the global feature group to separate the mid-sized van from all smaller and larger vehicle types. 
A general observation for the radio link groups is that the importance of higher-order signals (starting from $\Phi_3$) is reduced when the number of classes is increased. 
The latter is also confirmed by the subset-based analysis in the next section. 
More than 50\% of the total weight mass of the feature group $\Phi_2$ is responsible for the classification of mid-sized vehicles. 

%
%
\fig{}{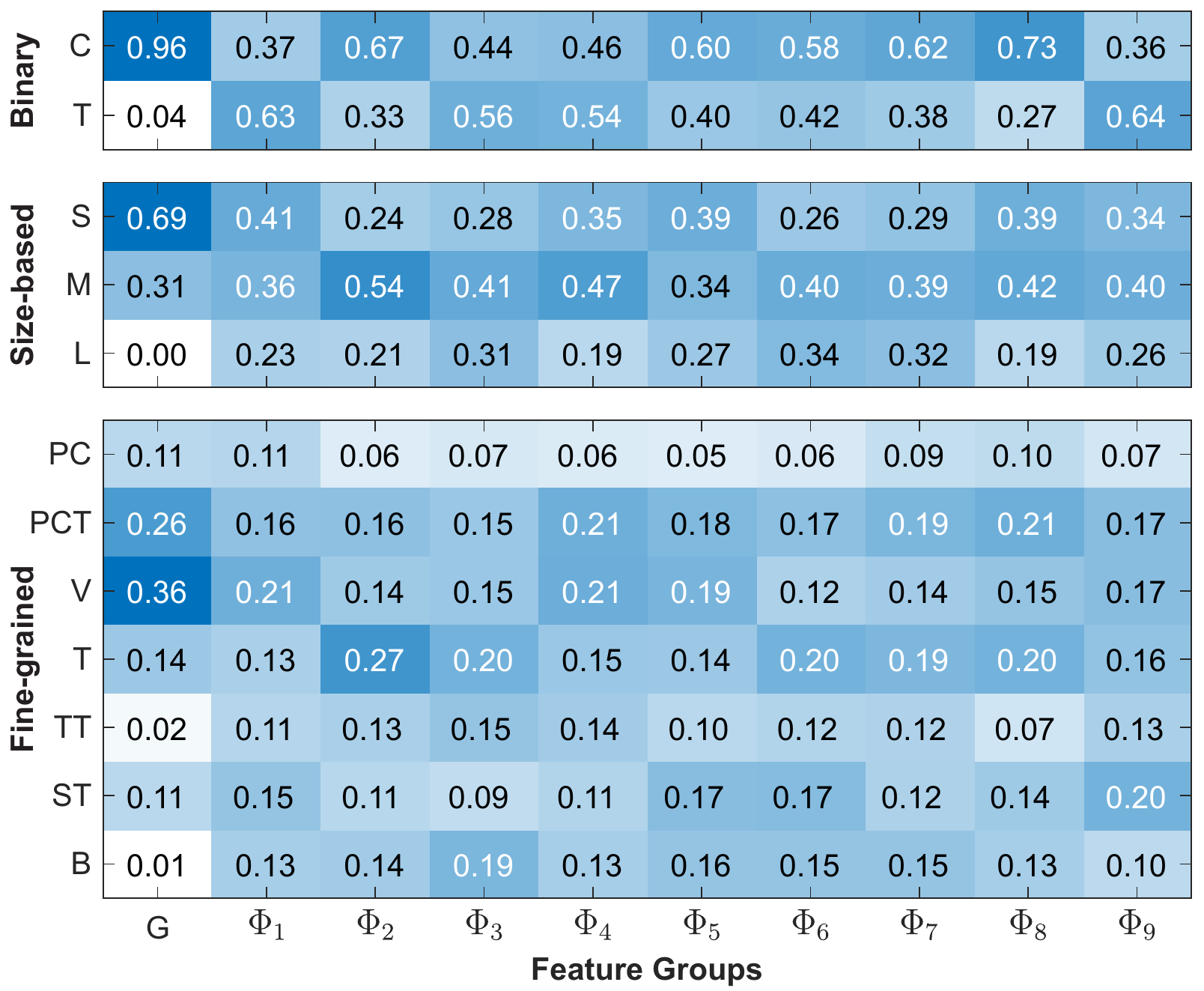}{Feature importance for \ac{SVM}-based classification based on link-wise weight aggregation. \emph{PC: Passenger car, PCT: Passenger car with trailer, V: Van, T: Truck, TT: Truck with trailer, ST: Semitruck, B: Bus.}}{fig:feature_importance}

\subsection{Subset-based Scalability Analysis}

It is obvious that the considered system configuration with nine different radio links is highly redundant. Therefore, it is now investigated how much the detection and classification performance is depending on the availability of multiple links and if a reduced system configuration would still be functional. Different aspects have to be remarked:
\begin{itemize}
	\item The \textbf{detection} performance is different for the individual links.
	\item Detecting \textbf{wrong-way drivers} is not possible if only a single pair of nodes is used since the speed estimation relies on analyzing the time differences of the attenuation phases of multiple signals.
\end{itemize}
%
%
%
%
%

\newcolumntype{D}{>{\centering\arraybackslash}m{0.46in}}
\newcommand\icon[1]{\includegraphics[width=0.15\columnwidth]{fig/subsets/#1}}

\begin{table*}[hb]
	\vspace{-0.5cm}
	\centering
	\caption{System Dimensioning Evaluation: Resulting Accuracy [\%] for Different Signal Subsets and Classification Taxonomies. Subsets: Binary (Two classes), Size-based (Three classes), Body style (Seven classes).}
	\begin{tabular}{cDDDDDDDDDD}
		
		\multirow{3}{0.7cm}{\textbf{Subset}}
		& \icon{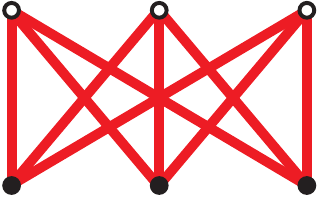} & \icon{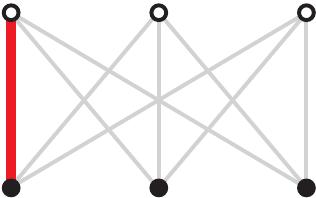} & \icon{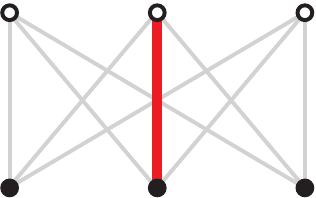} & \icon{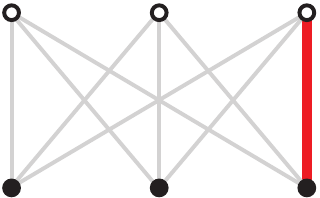} & \icon{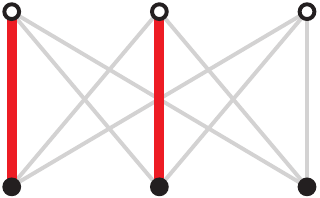} & \icon{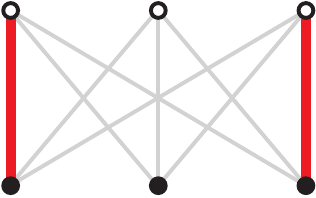} & \icon{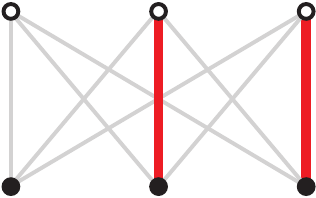} & \icon{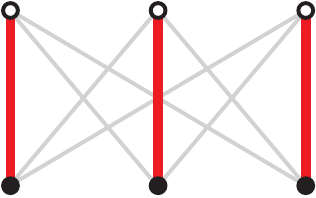} & \icon{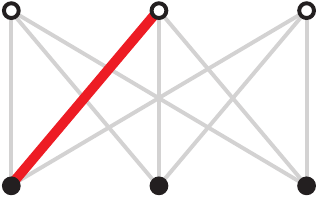} & \icon{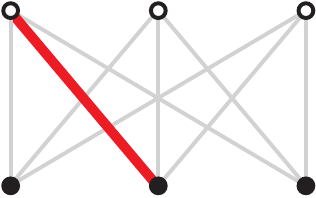}\\
		& \circled{A} & \circled{B} & \circled{C} & \circled{D} & \circled{E} & \circled{F} & \circled{G} & \circled{H} & \circled{I} & \circled{J}\\
		& $\Phi_{1-9}$ & $\Phi_{1}$ & $\Phi_{5}$ & $\Phi_{9}$ & $\Phi_{1,5}$ & $\Phi_{1,9}$ &  $\Phi_{5,9}$ &  $\Phi_{1,5,9}$ &  $\Phi_{2}$ & $\Phi_{4}$ \\
		
		\midrule
		\textbf{Binary} & $99.08\pm0.60$ & $98.62\pm0.49$ & $98.43\pm0.93$ & $98.23\pm0.67$ & $98.93\pm0.61$ & $98.73\pm0.62$ & $98.50\pm0.70$ & $99.08\pm0.43$ & $98.00\pm0.82$ & $98.35\pm0.45$ \\
		\midrule
		\textbf{Size-based} & $\mathbf{95.69}\pm\mathbf{0.99}$ & $94.59\pm1.45$ & $94.81\pm1.30$ & $94.28\pm0.84$ & $95.66\pm1.95$ & $95.17\pm1.33$ & $95.44\pm1.11$ & $95.35\pm1.42$ & $94.66\pm1.11$ & $94.59\pm1.34$ \\
		\midrule
		\textbf{Body style} & $\mathbf{93.82}\pm\mathbf{0.67}$ & $92.23\pm1.83$ & $91.98\pm0.76$ & $90.65\pm1.22$ & $92.83\pm1.96$ & $92.64\pm1.37$ & $92.57\pm1.28$ & $92.86\pm1.50$ & $91.37\pm1.21$ & $91.15\pm1.75$
		\\
		\midrule
		\multirow{3}{0.7cm}{\textbf{Subset}}
		& \icon{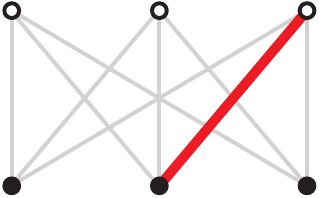} & \icon{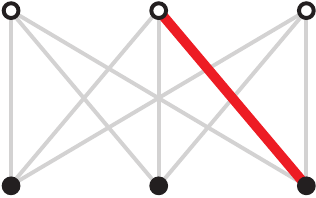} & \icon{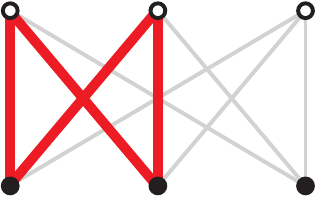} & \icon{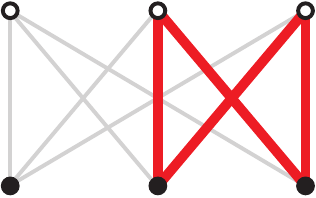} & \icon{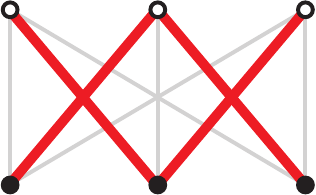} & \icon{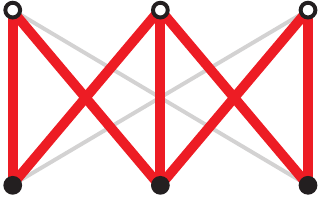} & \icon{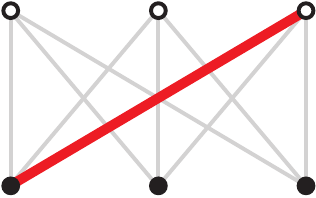} & \icon{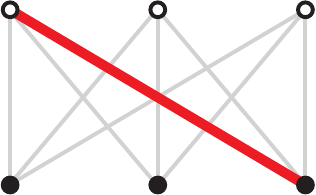} & \icon{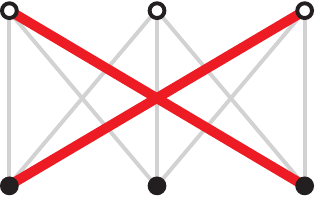} & \icon{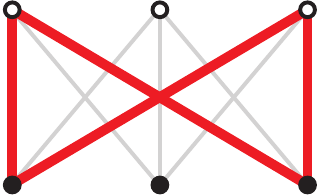}  \\
		
		& \circled{K} & \circled{L} & \circled{M} & \circled{N} & \circled{O} & \circled{P} & \circled{Q} & \circled{R} & \circled{S} & \circled{T}\\
		
		& $\Phi_{6}$ & $\Phi_{8}$ & $\Phi_{1,2,4,5}$ & $\Phi_{5,6,8,9}$ &  $\Phi_{2,4,6,8}$ & $\Phi_{1,2,4,6,8,9}$ &  $\Phi_{3}$ & $\Phi_{7}$ & $\Phi_{3,7,}$ & $\Phi_{1,3,7,9}$ \\

		\midrule
		\textbf{Binary} & $98.43\pm0.55$ & $97.89\pm0.67$ & $\mathbf{99.16}\pm\mathbf{0.64}$ & $98.89\pm0.58$ & $98.27\pm0.62$ & $98.96\pm0.64$ & $97.89\pm1.03$ & $98.24\pm0.94$ & $98.24\pm0.54$ & $99.04\pm0.35$ \\
		\midrule
		\textbf{Size-based} & $94.18\pm1.24$ & $93.28\pm1.30$ & $95.51\pm0.90$ & $95.54\pm0.87$ & $95.10\pm1.20$ & $95.55\pm1.49$ & $91.75\pm0.99$ & $93.01\pm0.89$ & $94.19\pm1.40$ & $95.44\pm1.47$ \\
		\midrule
		\textbf{Body style} & $90.57\pm1.02$ & $89.64\pm1.05$ & $93.25\pm0.80$ & $92.95\pm1.38$ & $92.23\pm1.73$ & $93.06\pm1.34$ & $87.81\pm1.26$ & $88.85\pm1.90$ & $90.41\pm1.61$ & $92.99\pm1.59$ \\

	\end{tabular}
	\label{tab:subsets}
\end{table*}

A summary of the resulting classification accuracies for different subset configurations is shown in Tab.~\ref{tab:subsets}. The applicability of using a reduced amount of radio links highly depends on the applied taxonomy and the implied classification granularity. Considering the standard deviations for each taxonomy, many of the configurations achieve a comparable performance level as the full system configuration. This observation shows that reduced system variants are in general feasible and might be considered as more cost-efficient alternative to full system deployments.
Binary classification works well with different subsets which contain at least two of the straight radio links. More than 99\% accuracy is achieved for the configurations \circled{M}, \circled{H}, and \circled{T}.
%
%
The same applies for the size-based classification task with three different vehicles types where different configurations make it possible to increase the cost efficiency of the proposed system by only deploying a reduced amount of sensor nodes. For example, configuration \circled{M} achieves a very good classification accuracy and allows to remove two nodes from the system.
%
%
However, for the body style classification task with seven different vehicle types, there is a significant loss in accuracy for most configurations that only consider a subset of the radio links. Still, some configurations (e.g.,\circled{M}) are plausible candidates for being considered in a reduced system deployment.

\subsection{Deployment of Classification Models to Real World \ac{IoT} Platforms} \label{sec:platforms}

In the following, we practically analyze dimensioning aspects for the real world application of the trained models on resource-constrained \ac{IoT} platforms. Tab.~\ref{tab:limits_iot_platforms} summarizes the key properties and resource limitations of the considered \acp{MCU}. For this purpose, we embed the platform-specific compilation toolchain into the data analysis process using the \ac{LIMITS} \cite{Sliwa/etal/2020c} framework which allows us to assess the real memory usage on the target platform.

%
%
%
%
%
\begin{table}[t]
	\centering
	\caption{Specifications of the Target \ac{IoT} Platforms.}
	\begin{tabular}{llll}
		
		\toprule
		\textbf{Parameter} & \textbf{\msp} & \textbf{\atmega} & \textbf{\esp} \\
		\midrule
		
		Model name & G2553 & ATmega328P & ESP-WROOM-32\\
		\acs{CPU} frequency & 16~MHz & 16~MHz & 240~MHz \\
		\acs{RAM} & 512~B & 2~kB & 532~kB \\
		Program memory & 16.32~kB & 32~kB & 4~MB \\
		\acs{IDE} & CCS & Arduino & ESP-IDF \\
		
		\bottomrule
		
	\end{tabular}
	\label{tab:limits_iot_platforms}
	\vspace{-0.3cm}
\end{table}

%
%
\fig{}{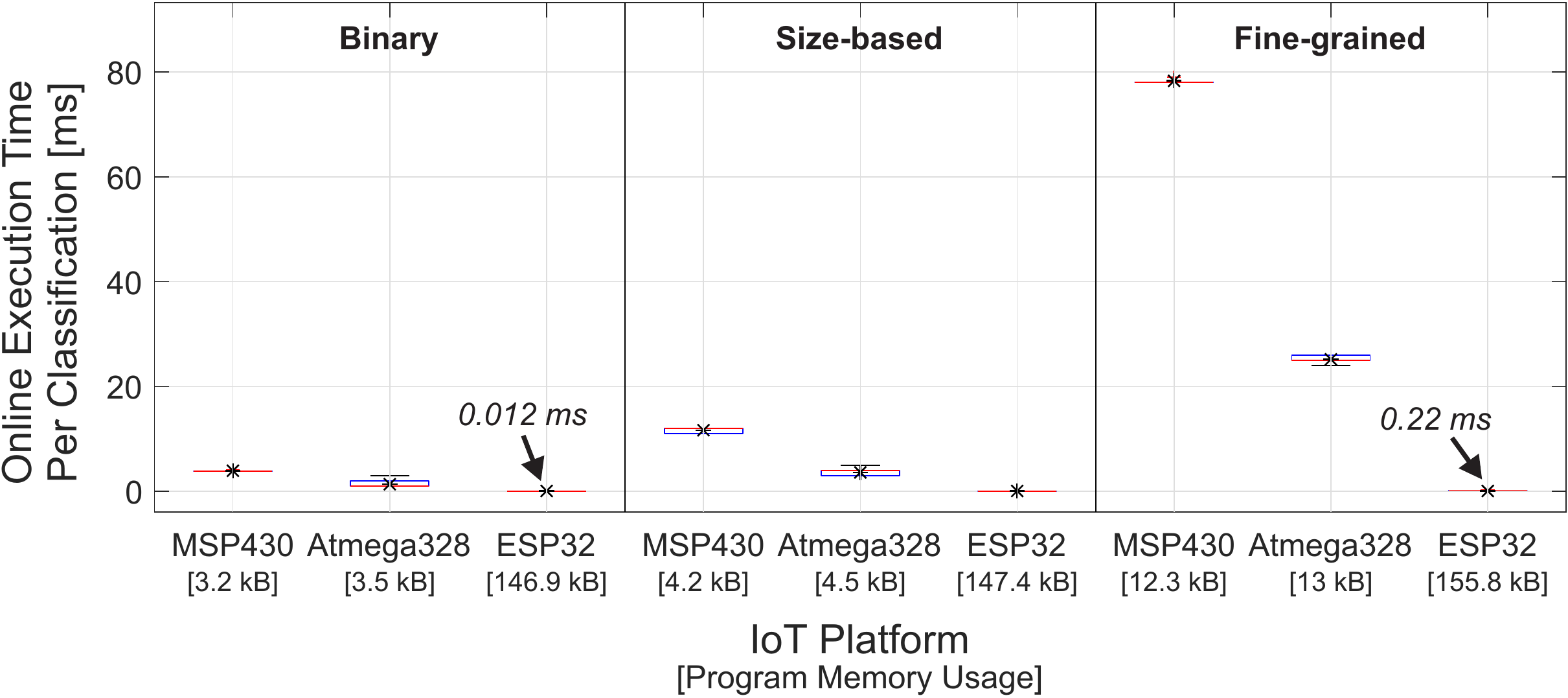}{Temporal effort for applying \ac{SVM}-based classification on different \ac{IoT} platforms.}{fig:platforms_time}

%
%
\begin{figure*}[] 
	\centering
	
	\subfigh{fig/rf_acc}{\sfw}{Model Performance}%
	\subfigh{fig/rf_msp}{\sfw}{\msp}%
	\subfigh{fig/rf_atmega}{\sfw}{\atmega}%
	\subfigh{fig/rf_esp}{\sfw}{\esp}%
	
	\caption{Parameter space exploration for determination of the platform-specific \emph{sweet spot} for the \ac{RF} prediction model for the body style classification task.}
	\vspace{-0.5cm}
	\label{fig:rf_sweet_spots}
\end{figure*}

Since the achievable classification interval is of crucial importance for the real world application, the temporal effort for execution the \ac{SVM} model on the considered target platforms is investigated. \ac{LIMITS} is used to generate the platform-specific \texttt{C++} code for the online execution of the machine learning models. After deploying the pre-trained models, the real world measurements are replayed directly on the different \ac{IoT} platforms and the temporal behavior is measured.
Fig.~\ref{fig:platforms_time} shows the resulting computation times per classification on different target \ac{IoT} platforms for the considered taxonomies. 
For the binary classification task, all platforms achieve a satisfying classification speed. 
Due to the applied \emph{one-vs-one} $Y$-class evaluation strategy for the \ac{SVM}, ${Y(Y-1)}/{2}$ comparisons are performed for a classification task with $Y$ classes. Therefore, the taxonomy has a dominant impact on the computation time.

As the considered linear \ac{SVM} is a \emph{parametric} model, the resulting model size is fixed and does not depend on the training data. However, for other models such as \ac{RF}, different variants have a more dominant impact on the resulting model size. Since the available program memory is one of the limiting factors for \acp{MCU}, it makes sense to determine the \emph{sweet spot} of a machine learning model for a given target platform. Hereby, a parameter space exploration is performed in order to find the most accurate model which just fits in the available memory.

Since the considered \ac{RF} model achieved the second best classification performance in Fig.~\ref{fig:results_box}, we further investigate the platform-specific dimensioning of different model variants.
Fig.~\ref{fig:rf_sweet_spots} shows the resulting accuracy and the required program memory size for different amounts of random trees and maximum depth values on the considered target \ac{IoT} platforms. 
The highly resource-constrained platforms \msp and \atmega are only able to implement a limited subset of the parameter space. As a result, the achievable accuracy is reduced by $\sim$1.5\% (for the \msp platform) respectively $\sim$0.5\% (for the \atmega platform) in comparison to the model parameterization with the highest accuracy (on the \esp platform). Although the \esp platform shows a significantly higher memory size due to additional mandatory libraries which are linked to the classification code within the compilation process, the large program memory allows to implement all parameterization variants.

\section{Conclusion} \label{sec:conclusion}

%
%
In this paper, we presented a novel traffic sensing system, which exploits attenuation patterns of wireless signals as \emph{radio fingerprints} for accurate vehicle detection and classification.
%
%
In contrast to existing solutions, the proposed system is able to provide a favorable set of system properties. It is accurate, privacy-preserving, cost-efficient and robust against typical disturbance factors.
In a comprehensive performance evaluation, the classification accuracy of different machine learning models was analyzed for multiple multi-class granularities. Using a linear \ac{SVM}, a binary classification accuracy of more than 99\% can be achieved.
%
%
In a first feasibility study, it was also shown that the radio-based approach can generally be applied for detecting wrong-way drivers based on vehicle speed estimations. Therefore, system deployments at motorway access roads could potentially contribute to increasing the highway road safety.
%
%
In future work, we want to investigate the applicability of different communication technologies (e.g., usage of highly-directed \ac{mmWave} beams) for radio-based vehicle classification. Moreover, we will evaluate the integration of additional sensors (e.g., \acp{IMU}) into the proposed system and analyze methods for multi lane detection and classification. In addition, we will further analyze the speed estimation accuracy of the proposed system based on dedicated measurements using \ac{RADAR} as a ground truth.

\section*{Acknowledgment}

\footnotesize
Part of the work on this paper has been supported by Deutsche Forschungsgemeinschaft (DFG) within the Collaborative Research Center SFB 876 ``Providing Information by Resource-Constrained Analysis'', projects B4. Parts of this research have been funded by the Federal Ministry of Education and Research of Germany as part of the competence center for machine learning ML2R (01$|$S18038B). The authors thank Dennis Dorn and S-tec GmbH for their support with performing the real world measurements.

\begin{IEEEbiography}[{\includegraphics[width=1in,height=1.25in,clip,keepaspectratio]{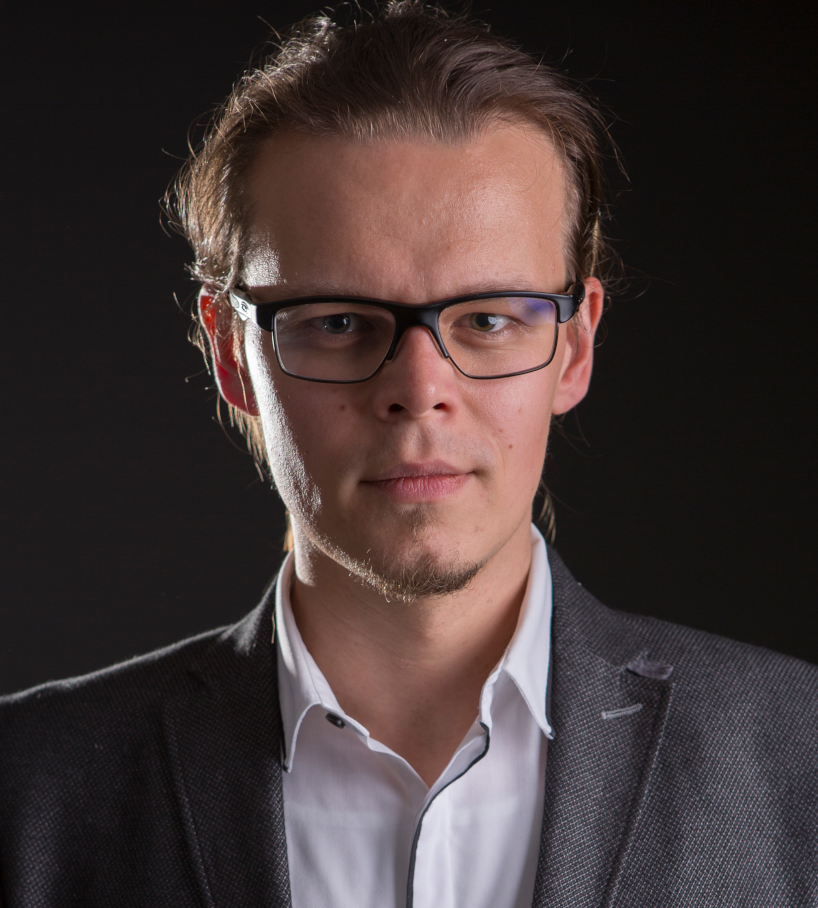}}]
	{Benjamin Sliwa}
	(S'16) received the M.Sc. degree from TU Dortmund University, Dortmund, Germany, in 2016. He is currently a Research Assistant with the Communication Networks Institute, Faculty of Electrical Engineering and Information Technology, TU Dortmund University. He is working on the Project "Analysis and Communication for Dynamic Traffic Prognosis" of the Collaborative Research Center SFB 876. His research interests include predictive and context-aware optimizations for decision processes in vehicular communication systems. Benjamin Sliwa has been recognized with a Best Student Paper Award at IEEE VTC-Spring 2018 and the 2018 IEEE Transportation Electronics Student Fellowship "For Outstanding Student Research Contributions to Machine Learning in Vehicular Communications and Intelligent Transportation Systems".
\end{IEEEbiography}
\begin{IEEEbiography}
[{\includegraphics[width=1in,height=1.25in,clip,keepaspectratio]{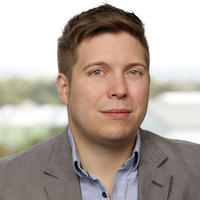}}]
{Nico Piatkowski}
received the Dipl.-Inf. and Dr.~rer.~nat. degrees from TU Dortmund University, Dortmund, Germany. He is currently a senior researcher at the Media Engineering Group of the Fraunhofer Institute for Intelligent Analysis and Information Systems IAIS in Sankt Augustin, Germany. Before that, he was a researcher at the DFG Collaborative Research Center SFB 876 on ``Providing Information by Resource-Constrained Data Analysis'' and later at the Competence Center Machine Learning Rhine-Ruhr (ML2R). Nico contributed to various research projects on the foundations of machine learning under resource constraints, spatio-temporal modelling, and probabilistic inference. His current research interests include deep latent variable models with discrete state spaces, high-dimensional spatio-temporal generative models, and regularized learning. Nico Piatkowski received ECML-PKDD's awards for being an outstanding program committee member and for his work on ``Spatio-Temporal Random Fields'', as well as TU Dortmund's dissertation award for his work on ``Exponential Families on Resource-Constrained Systems''. 
\end{IEEEbiography}
\begin{IEEEbiography}
[{\includegraphics[width=1in,height=1.25in,clip,keepaspectratio]{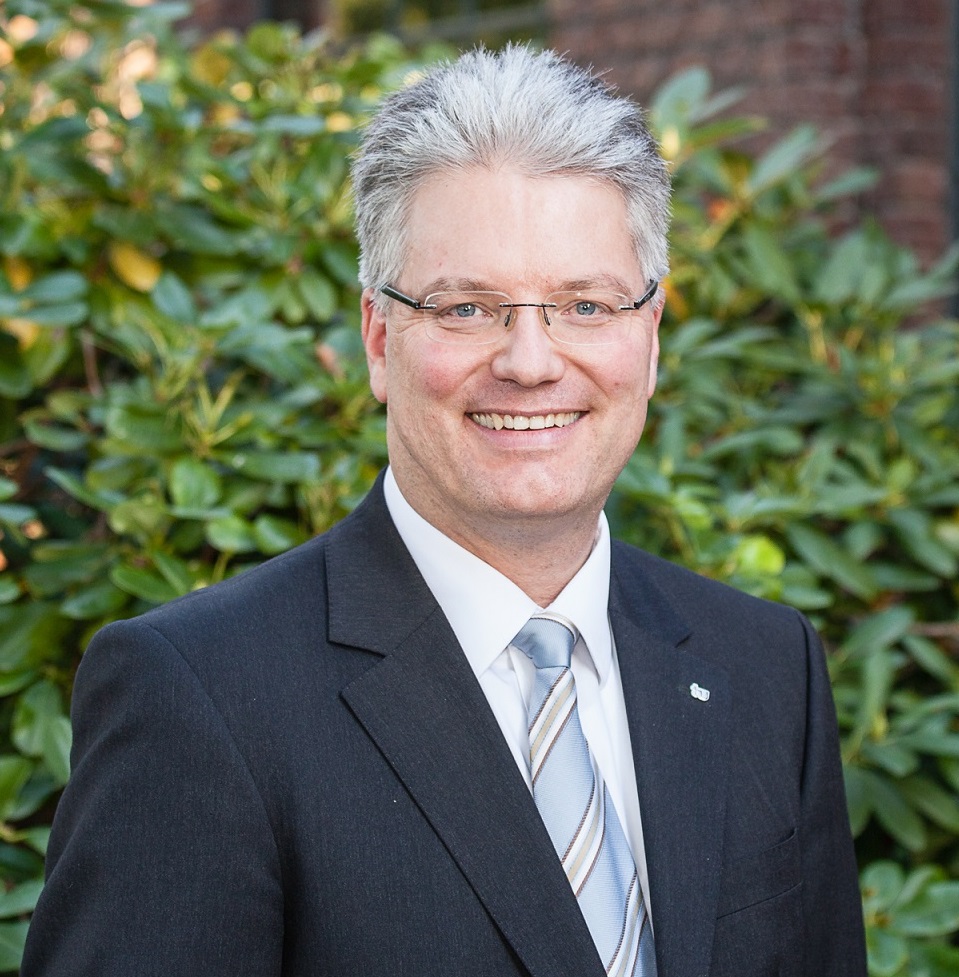}}]
{Christian Wietfeld}
(M’05–SM’12) received the Dipl.-Ing. and Dr.-Ing. degrees from RWTH Aachen University, Aachen, Germany.  He is currently a Full Professor of communication networks and the Head of the Communication Networks Institute, TU Dortmund University, Dortmund, Germany. For more than 20 years, he has been a coordinator of and a contributor to large-scale research projects on Internet-based mobile communication systems in academia (RWTH Aachen ‘92-’97, TU Dortmund since ‘05) and industry (Siemens AG ’97-’05). His current research interests include the design and performance evaluation of communication networks for cyber–physical systems in energy, transport, robotics, and emergency response.  He is the author of over 200 peer-reviewed papers and holds several patents. Dr. Wietfeld is a Co-Founder of the IEEE Global Communications Conference Workshop on Wireless Networking for Unmanned Autonomous Vehicles and member of the Technical Editor Board of the IEEE Wireless Communication Magazine. In addition to several best paper awards, he received an Outstanding Contribution award of ITU-T for his work on the standardization of next-generation mobile network architectures.

\end{IEEEbiography}

\bibliographystyle{IEEEtran}
\bibliography{Bibliography}

\end{document}